\RequirePackage{silence}
\documentclass[10pt,twocolumn,aps,prb]{revtex4-2}
\usepackage[english]{babel} 
\usepackage[usenames,dvipsnames]{color}  
\usepackage{graphicx} 
\usepackage{bm} 
\usepackage{amsmath}
\usepackage{amsthm} 
\usepackage{amssymb} 
\usepackage{times} 
\usepackage{mathrsfs}
\usepackage{xr-hyper}
\usepackage{newpxtext}
\usepackage[hypertexnames=false,pdftex,colorlinks=true,urlcolor=blue,
            linkcolor=Blue,citecolor=RedViolet]{hyperref} 
\newcommand{\upcite}[1]{\textsuperscript{\citenum{#1}}} 
\begin{document}
\title{Williamson theorem in classical, quantum, and statistical physics}
\author{F. Nicacio}
\email{nicacio@if.ufrj.br} 
\affiliation{Instituto de F\'isica, Universidade Federal do Rio de Janeiro, 
             21941-972, RJ, Brazil}
\affiliation{ 
              Universit\"at Wien, NuHAG, Fakult\"at f\"ur Mathematik, 
              A-1090 Wien, Austria.                                   }             

\date{\today}

\begin{abstract} 
In this work we present (and encourage the use of) 
the Williamson theorem and its consequences in several contexts in physics. 
We demonstrate this theorem using only basic concepts 
of linear algebra and symplectic matrices.      
As an immediate application in the context of small oscillations, 
we show that applying this theorem reveals the normal-mode 
coordinates and frequencies of the system 
in the Hamiltonian scenario. 
A modest introduction of the symplectic formalism in quantum mechanics is presented, 
useing the theorem to study quantum normal modes and canonical distributions of 
thermodynamically stable systems described by quadratic Hamiltonians. 
As a last example, a more advanced topic concerning uncertainty relations 
is developed to show once more its utility in a distinct and modern perspective.
\end{abstract}

\maketitle 

\section{Introduction}                                         
The main advantage of the Hamiltonian formalism in classical mechanics 
is the symmetry of the equations of motion with respect to 
position and momentum coordinates, 
which naturally embody the symplectic structure of
the phase space\upcite{arnold,landau1,goldstein,lemos,arnold}.
The same structure is also present in quantum mechanics 
through position and momentum operators of the systems\upcite{littlejohn1986,gossonbook2006}, 
which in either classical or quantum physics is the arena for the Williamson theorem 
that  describes a diagonalization procedure suitable to the symplectic scenario. 
%
%
%
%
Just as diagonalizing a matrix 
in  Euclidean space determines invariant quantities 
(eigenvalues and eigenvectors), applying the Williamson theorem 
reveals various properties of symplectic invariance. 
%

%
The initial part of this paper, Section \ref{Sec:wt}, 
intro\-du\-ces the mathematical notation and then presents 
the Williamson theorem, which is proved in the 
Supplementary Material\upcite{SupMatDem} 
using only basic concepts in linear algebra. 
%
%

The central application is the study of small oscillations 
in the context of Hamiltonian dynamics, 
which is performed by the diagonalization of a positive-definite quadratic form 
through the use of the theorem. 
%
%
To present this study, Sec.\ref{Sec:HM} reviews Hamiltonian mechanics and then treats 
quadratic Hamiltonians using the theorem. 
%
The standard method of dealing with the problem of small oscillations 
(the simultaneous diagonalization of the kinetic and potential terms of 
a Lagrangian function\upcite{arnold,landau1,goldstein,lemos}) 
is compared with the Hamiltonian results in the Supplementary Material\upcite{SupMatLag}. 
%
%
The advantages of the Williamson theorem become clear in this context: 
a change of coordinates in phase space reveals the normal modes 
and the eigenfrequencies of the system. 
%
%
 
In Section \ref{Sec:QM}, 
initial concepts of the symplectic formalism in quantum mechanics 
are described that allow the theorem to be used 
to study small oscillations in quantum systems. 
Because creation-annihilation operators\upcite{sakurai,ballentine,cohen} are often used 
in study of oscillations, these operators are placed in a (complex) symplectic scenario, 
suitable to the application of the theorem.

%

The previous applications lead immediately to the use 
of the theorem to study the canonical equilibrium ensemble 
of statistical physics. 
In Section \ref{Sec:SM}, 
the equilibrium state and the partition function\upcite{landau2,huang,pathria}
associated with a generic quadratic Hamiltonian 
are determined for thermodynamically stable systems, 
where the normal-mode frequencies play the fundamental role, 
showing that all the thermodynamical properties of the system are 
symplectically invariant.

Crossing the frontier towards modern research,
Section \ref{Sec:UP} contains a pedagogical
derivation for the Ro\-ber\-tson-Sch\-rödinger uncertainty relation, which is 
a generalization of the Heisenberg principle\upcite{sakurai,ballentine,cohen}. 
The application of the theorem reveals 
invariant properties common to all physical states. 
This content is inspired by the results in Ref.\onlinecite{simon1994},
probably the first paper in physics introducing 
the theorem in the sense presented here. 

Section \ref{Sec:FR} concludes by presenting comments on generalizations 
of the theorem and references to modern applications. 
The idea behind this manuscript is to bring it to classroom, 
showing how standard problems in physics courses can be treated 
using this simple and unified perspective.

%
Physically motivated examples are presented in 
the Supplementary Material\upcite{SupMatEx}.   

\subsection*{A starting example:}%
Consider a system with one degree of freedom described by the Hamiltonian 
\begin{equation}
H(q,p) = \frac{a}{4} ( q/q_0+p/p_0)^2 + \frac{b}{4} (q/q_0-p/p_0)^2, 
%
\end{equation}
where $q$ is the generalized coordinate; 
$p$ the canonically conjugated momentum; and   
$a$, $b$, $q_0$, and $p_0$ are real constants.  
Without loss of generality, one can choose $q_0 p_0 = 1$, 
which is nothing but a choice of units. 
If $a = b$, the Hamiltonian describes a harmonic oscillator, 
{\it i.e.}, $H(q,p) = \frac{a}{2} [(q/q_0)^2 + (p/p_0)^2]$. 
Are there other possibilities for which the original Hamiltonian 
describes harmonic motion? 
Basically, this is the question posed in this work. 

The Hamilton equations of motion\upcite{arnold,landau1,goldstein,lemos} 
for the original Hamiltonian are
\begin{equation}
\begin{aligned}
\dot q &= \frac{\partial H}{\partial p} = 
\tfrac{1}{2}(a-b) q + \tfrac{1}{2}(a+b) p/p_0^2, \\
\dot p &= - \frac{\partial H}{\partial q} = 
-\tfrac{1}{2}(a+b) q/q_0^2 + \tfrac{1}{2}(b-a) p, 
\end{aligned}
\end{equation}
which can be rearranged as the vector equation 
$\dot x = {\bf A} x$ for  
\begin{equation}
x := \left(\begin{array}{c}
     q \\ p
     \end{array}\right), \,\,\,
{\bf A} := 
          \left(\begin{array}{cc}
                 \displaystyle \frac{a-b}{2}  & 
                 \displaystyle \frac{a+b}{2p_0^2} \\
                 \displaystyle -\frac{a+b}{2 q_0^2}  & 
                 \displaystyle \frac{b-a}{2}
                \end{array}\right).
\end{equation}

By a suitable linear (canonical) change of coordinates, 
$x = {\mathsf S} x'$, 
the two coupled equations become $\dot x' = {\bf A}' x'$ with 
\begin{equation}
\begin{aligned}
{\bf A}' &:= {\mathsf S}^{-1} {\bf A} {\mathsf S} = 
\begin{pmatrix}
0 & \sqrt{ab}\\
-\sqrt{ab} & 0
\end{pmatrix}, \\
\mathsf S &= 
\begin{pmatrix}
q_0 & 0\\
0 & p_0
\end{pmatrix}
\begin{pmatrix}
\frac{1}{\sqrt{2}} & -\frac{1}{\sqrt{2}}\\
\frac{1}{\sqrt{2}} & \frac{1}{\sqrt{2}}
\end{pmatrix}
\begin{pmatrix}
\sqrt[4]{\frac{b}{a}} & 0\\
0 &  \sqrt[4]{\frac{a}{b}}
\end{pmatrix}
,  
\end{aligned} 
\end{equation}
and the equations of the motion for the new pair 
of coordinates are 
\begin{equation}
\dot q' = \sqrt{ab} \, p', \,\,\,  
\dot p' = -\sqrt{ab} \, q'.
\end{equation}
These are the Hamilton equations for the Hamiltonian 
$H' = \tfrac{\sqrt{ab}}{2} (p'^2 + q'^2)$, corresponding 
to a harmonic oscillator if $ab >0$. 
Note that $\ddot q' = \sqrt{ab} \, \dot p'$, 
and thus $\ddot q' = - ab\, q'$. 

The matrix $\mathsf S$ performs an 
anti-diagonalization of $\bf A$.   
$\mathsf S$ is a {\it symplectic matrix} and $\sqrt{ab}$ 
is a {\it symplectic eigenvalue}. 
These concepts will be defined soon; 
for now it is enough to state that every $2 \times 2$ real matrix 
with unity determinant is symplectic 
and that the symplectic eigenvalue is not equal to
an ordinary (Euclidean) eigenvalue.  

The relation between the Hamiltonian and matrix $\mathsf S$ is established
when considering the Hessian
\begin{equation}
{\mathbf H} :=  
\left(\begin{array}{cc}
\frac{\partial^2 \!H}{\partial q^2} & 
\frac{\partial^2 \!H}{\partial q \partial p}\\
\frac{\partial^2 \!H}{\partial p \partial q} & 
\frac{\partial^2\! H}{\partial p^2}
\end{array}\right)
= 
\left(\begin{array}{cc}
                 \displaystyle \frac{a+b}{2 q_0^2}  & 
                 \displaystyle \frac{a-b}{2} \\
                 \displaystyle \frac{a-b}{2}  & 
                 \displaystyle \frac{a+b}{2p_0^2}
                \end{array}\right), 
\end{equation}
which is such that $\mathsf S^\top {\bf H} \mathsf S = 
{\rm Diag}(\sqrt{ab},\sqrt{ab}) =: {\bf H}'$, 
where $\mathsf S^\top$ is the transpose of $\mathsf S$. 
Here the matrix $\mathsf S$ performs a {\it symplectic diagonalization} 
of ${\bf H}$, which is not a coincidence, but rather a consequence 
of the symplectic structure of phase space manifested through the 
identity 
\begin{equation}
{\bf A} = {\mathsf J}{\bf H}, \,\,\, {\text{where}} \,\,\,
{\mathsf J} := \left(
\begin{array}{cc}
 0 & 1\\
-1 & 0
\end{array}\right). 
\end{equation}
Noteworthy: ${\bf A}' = {\mathsf J}{\bf H}'$, 
where ${\bf H}'$ is the Hessian of $H'(q',p')$. 

Diagonalizing the matrix $\bf A$ reveals that 
its eigenvalues are $\pm i \sqrt{ab}$, which are complex if $ab>0$. 
The equations of motion could be decoupled using the 
(complex) coordinates $w := {\bf B} x$, 
where ${\bf A}'' = {\bf B}{\bf A}{\bf B}^{-1}$ is diagonal. 
%
However, all the phase-space properties would be lost; 
for instance, it would be impossible to attain 
a real Hamiltonian function for the decoupled degrees of freedom. 
The great advantage of the symplectic change of coordinates 
$x= \mathsf S x'$ is that it rewrites the dynamics of the original 
system as a mechanically equivalent system, 
preserving all the structure and symmetry of phase space.  

The Lagrangian function for the same system is obtained by 
the Legendre transformation\upcite{arnold,landau1,goldstein,lemos} 
of the original Hamiltonian, 
\begin{equation}
L(q,\dot q) = p(q,\dot q) \dot q - H(q,p(q,\dot q)) = 
\frac{(\dot q - a q)(\dot q + b q)}{(a+b) q_0^2}, 
\end{equation}
where the function $p(q,\dot q)$  
was obtained from equation 
$\dot q = \partial H/\partial p$ 
to be $p = [2 \dot q + (b-a) q]/[(a+b)q_0^2]$. 
Using the Euler-Lagrange equation\upcite{arnold,landau1,goldstein,lemos},  
the generalized coordinate satisfies 
$\ddot q + (ab)\, q = 0$, 
which is the same requirement as obtained for $q'$ in the Hamiltonian scenario. 
Up to this point, the Lagrangian treatment seems to be simpler and 
straightforward. 

However, the Heisenberg equations\upcite{sakurai,ballentine,cohen} 
for the dynamics governed by the quantization of the original Hamiltonian, 
\begin{equation}
\hat H := H(\hat q, \hat p) = 
\frac{a}{4} ( \hat q/q_0+\hat p/p_0)^2 + 
\frac{b}{4} (\hat q/q_0-\hat p/p_0)^2,
\end{equation}
are given by 
\begin{equation}
\begin{aligned}
\frac{d \hat q}{dt} &= \tfrac{i}{\hbar}
[ \hat H, \hat q ] = 
\tfrac{1}{2}(a-b) \hat q + \tfrac{1}{2}(a+b) \hat p/p_0^2, \\
\frac{d \hat p}{dt} &= \tfrac{i}{\hbar}
[ \hat H, \hat p ] =
-\tfrac{1}{2}(a+b) \hat q/q_0^2 + \tfrac{1}{2}(b-a) \hat p, 
\end{aligned}
\end{equation}
which are the same as the classical ones if one
replaces $q \mapsto \hat q$ and $p \mapsto \hat p$. 
It is thus possible to apply the same linear canonical transformation at 
the operator level, attaining  the equivalence with a Hamiltonian system of 
quantum oscillators under the same condition $ab > 0$. 
This compatibility of classical and quantum scenarios clearly 
constitutes a huge advantage over the Lagrangian description. 

An immediate but not obvious question is to what extent 
the above symplectic procedure can be applied to more complex 
(classical or quantum) systems. 
The answer will be given by the Williamson theorem, which will 
provide conditions for a Hamiltonian system to behave like a set of 
harmonic oscillators. 

Another introductory example can be found 
in Supplementary Material\upcite{SupMatEx2},
where the Lagrangian treatment of small oscillations is performed and 
compared with the symplectic diagonalization scheme 
for a physical interesting problem, namely,   
the dynamics of two interacting trapped ions. %

\section{Williamson Theorem}                                    \label{Sec:wt}     
The question addressed by the Williamson theorem is the diagonalization of positive 
definite matrices through symplectic matrices. 
Before the presentation of the theorem, 
some basic concepts concerning these kinds of matrices 
and some linear algebra will be reviewed. 

A vector $v \in \mathbb R^n$ is a column of $n$ real components $v_i$, with $i=1,...,n$ 
and its transposition is the line vector $v^\top := (v_1,...,v_n)$. 
The scalar product between $u,v \in \mathbb R^n$ is defined by  
$u \cdot v := u^\top v = \sum_{i = 1 }^n u_i v_i \in \mathbb R$.  
For two complex vectors $z,w \in \mathbb C^n$, 
their scalar product is $z^\dag w := \sum_{i = 1 }^n z_i^\ast w_i \in \mathbb C$, where $z^\dag := (z_1^\ast,...,z_n^\ast)$.  
The set of all $n \times n$ complex square matrices 
is denoted by ${\rm M}(n)$, and for real matrices
the notation ${\rm M}(n, \mathbb R)$ will be used. 
Note that ${\rm M}(n, \mathbb R) \subset {\rm M}(n)$.  
The identity and null matrices in ${\rm M}(n)$ 
are respectively denoted by
${\bf I}_n$ and ${\bf 0}_n$. 

Two matrices ${\bf A},{\bf B} \in {\rm M}(n)$ 
are said {\it similar}, 
if there exists an invertible ${\bf C} \in {\rm M}(n)$ 
such that ${\bf A} = {\bf C}{\bf B} {\bf C}^{-1}$. 
This relation corresponds to a change of basis in linear algebra, 
{\it i.e.}, $w = {\bf B} z$ is equivalent to ${\bf C}w = {\bf A} {\bf C}z$, 
for $z \in \mathbb C^n$.  
From this point of view, a similarity is related to structures 
of the transformation that are common to any basis of the space. 
%
The matrices ${\bf A}$ and ${\bf B}$ in this case share the 
same spectrum; that is, they have the same eigenvalues, 
since $\det({\bf A} - \lambda {\bf I}_n) = 
\det({\bf B} - \lambda {\bf I}_n)$. 
In this perspective, eigenvalues are invariant 
under a similarity relation, 
while eigenvectors are covariant; that is,  
if $z$ is an eigenvector of ${\bf B}$, 
then ${\bf C}z$ is an eigenvector of ${\bf A}$.
%
%
A diagonalizable matrix is the one that is similar to a diagonal matrix and
the {spectral theorem}\upcite{horn2013} sets a necessary and sufficient condition 
for it: a matrix ${\bf A} \in {\rm M}(n)$ is {\it normal}  
({\it i.e.}, ${\bf A}\!^\dag{\bf A} = {\bf A}{\bf A}\!^\dag$) 
if and only if it is unitarily similar to 
a diagonal matrix, which contains the eigenvalues of $\bf A$. 
For the similarity relation above, 
this means that ${\bf A}$ is normal if and only if 
there is $\bf C$ satisfying ${\bf C}^\dag = {\bf C}^{-1}$ 
such that $\bf B$ is diagonal.


Unitary matrices, which include either complex or real orthogonal matrices, 
are isometries of the Euclidean space, 
which means that they preserve the scalar product (or the ``distance''):  
${w^\dagger z = ({\bf U} w)^\dagger}{\bf U}z$,  
since ${\bf U}^\dagger{\bf U} = {\bf I}_{n}$ 
for ${\bf U} \in {\rm M}(n)$ and $w,z \in {\mathbb C}^{n}$. 
%
%
Whenever a diagonalization of a matrix is performed through  
either an orthogonal or a unitary similarity relation, 
which is the common sense for a diagonalization (through the spectral theorem), 
it will be called an {\it Euclidean diagonalization} and
the eigenvalues as the Euclidean eigenvalues. 
This nomenclature emphasizes the difference from another kind 
of diagonalization performed in the Williansom theorem, 
which will be a symplectic diagonalization. 

A weaker relation than similarity, but no less important here,  
is called {\it congruence}. 
Two matrices ${\bf A},{\bf B} \in {\rm M}(n)$ 
are said to be {\it congruent} if there exists a invertible 
${\bf C} \in {\rm M}(n)$ 
such that ${\bf A} = {\bf C}{\bf B} {\bf C}^{\dagger}$.  
Now, neither the spectrum nor the eigenvectors play a privileged role; 
however the inertia\upcite{Note:InerDef}
of $\bf A$ and $\bf B$ will be the same if and only if 
these matrices are Hermitian.   
This invariance property is known as   
{\it Sylvester's law of Inertia}\upcite{Note:InerLaw}, 
a kind of ``spectral theorem'' for congruence relations.  
When matrix $\bf C$ is unitary or orthogonal, 
the congruence ${\bf A} = {\bf C}{\bf B} {\bf C}^{\dagger}$ 
is also a similarity.  

A matrix ${\bf A} \in {\rm M}(n)$ 
is said {\it positive-definite}, denoted by ${\bf A} > 0$, 
if $w^\dag {\bf A} w > 0$, $\forall w \ne 0$ and 
$w \in \mathbb C^n$;  
if $\bf A$ is Hermitian, ${\bf A} = {\bf A}^\dag$, 
the last statement is equivalent to saying that all eigenvalues 
of $\bf A$ are real and positive\upcite{horn2013}.  
Consequently, 
all Hermitian positive-definite matrices are invertible, 
since $\det {\bf A} > 0$.   
For ${\bf A}^\dag = {\bf A} > 0$,  
the unique matrix $\sqrt{\bf A} \in {\rm M}(n)$ satisfying 
\begin{equation}
(\sqrt{\bf A} )^2 = {\bf A} 
\,\,\, \text{and} \,\,\,   
\sqrt{\bf A}\,^\dag = \sqrt{\bf A} > 0
\end{equation}
is the {\it positive square-root}\upcite{horn2013} of $\bf A$. 
If the eigenvalues of a Hermitian matrix 
${\bf A} \in {\rm M}(n)$  
are non-negative (they can be either positive or zero), 
the matrix is {\it positive semi-definite} (denoted by ${\bf A} \ge 0$) 
and is equivalent to $w^\dag {\bf A} w \ge 0$, $\forall w \ne 0$. 
A trivial corollary of the Sylvester law
relates positivity and congruences: 
Let ${\bf A}, {\bf B}, {\bf C} \in {\rm M}(n)$, 
such that 
${\bf A} = {\bf C}{\bf B} {\bf C}^{\dagger}$ with $\det {\bf C} \ne 0$;
for $w \in \mathbb C^n$, 
$w^\dag {\bf A}w = ({\bf C}^\dag w)^\dag {\bf B} {\bf C}^\dag w$, 
thus 
${\bf A} \ge 0$ (resp. ${\bf A} > 0$) if and only if 
${\bf B} \ge 0$ (resp. ${\bf B} > 0$).



A matrix ${\bf M} \in {\rm M}(2n)$ 
can be written as a block matrix  
when portioned by smaller matrices\upcite{horn2013}:    
\begin{equation}
{\bf M} =  
\begin{pmatrix}
           {\bf A}   & {\bf B} \\
           {\bf C}   & {\bf D} 
\end{pmatrix}, \,\,\, 
{\bf A},{\bf B},{\bf C},{\bf D} \in {\rm M}(n),  
\end{equation}
where ${\bf A}_{ij} = {\bf M}_{ij}$ for $i,j \le n$, 
${\bf B}_{ij} = {\bf M}_{ij}$ for $i \le n$ and $n+1 \le j \le 2n$, 
{etc}. 
As a compact and useful notation, the {\it direct sum}\upcite{horn2013} 
${\bf A} \oplus {\bf D}$ is a block-diagonal matrix, {\it i.e.}, 
the above matrix $\bf M$ with ${\bf B} ={\bf C} = {\bf 0}_{n}$. 
The determinant of a block matrix can be expressed in terms 
of its blocks\upcite{horn2013,silvester},  
for instance, 
$
\det{\bf M} = \det{\bf D} \det({\bf A}-{\bf B}{\bf D}^{-1}{\bf C})
$, 
if ${\bf D}$ is nonsingular. 
If in addition 
$[{\bf C},{\bf D}] = 0$, then 
$\det{\bf M} = \det({\bf AD}-{\bf BC})$.
All the above properties and formulas can be generalized  
for nonsquare blocks, different partitions, or even 
singular blocks\upcite{horn2013}. 

A {\it symplectic matrix} ${\sf S} \in {\rm M}(2n,\mathbb R)$ 
is defined by the rule 
\begin{equation} \label{Eq:SympCond}
{\sf S}^\top {\sf J} {\sf S} = {\sf J}, 
\,\,\, \text {where} \,\,\,  
{\sf J} = 
          \begin{pmatrix}
          {\bf 0}_n   & {\bf I}_n \\
          - {\bf I}_n & {\bf 0}_n 
          \end{pmatrix}               \in {\rm M}(2n,\mathbb R). 
\end{equation}
The matrix $\sf J$ is such that ${\sf J}^2 = - {\bf I}_{2n}$, 
thus ${\sf J}^{-1} = - {\sf J} = {\sf J}^\top$,
and is itself a symplectic matrix with $\det {\mathsf J} = 1$. 
Taking the transposition of ${\sf S}$, 
one shows that ${\sf S}^\top$ is also symplectic 
and that the condition ${\sf S} {\sf J} {\sf S}^\top = {\sf J}$ 
is equivalent to Eq.~(\ref{Eq:SympCond}).  
The determinant of a symplectic matrix, 
from the definition, is such that 
$\det{\sf S}^2 = 1$. 
Consequently, every symplectic matrix is invertible 
and the inverse is 
${\sf S}^{-1} = {\sf J}^\top {\sf S}^\top {\sf J}$ by Eq.~(\ref{Eq:SympCond}). 
Finally, the set of symplectic matrices forms the group 
\begin{equation}
{\rm Sp}(2n,\mathbb R) := \left\{ {\sf S} \in {\rm M}(2n,\mathbb R) \, | 
                             {\sf S}^\top {\sf J} {\sf S} ={\sf J}     \right\},
\end{equation}
since 
${\bf I}_{2n} \in {\rm Sp}(2n,\mathbb R)$, 
${\sf S}^{-1} \in {\rm Sp}(2n,\mathbb R)$ 
if ${\sf S} \in {\rm Sp}(2n,\mathbb R)$, 
and
${\sf S}_1{\sf S}_2 \in {\rm Sp}(2n,\mathbb R)$ 
if
${\sf S}_1, {\sf S}_2 \in {\rm Sp}(2n,\mathbb R)$.  
It is not difficult to show that condition Eq.~(\ref{Eq:SympCond}) 
reduces to $\det {\mathsf S} = 1$ for $n = 1$; in other words, 
every $2 \times 2$ real matrix with determinant one 
is a symplectic matrix. 
Every matrix ${\mathsf S} \in {\rm Sp}(2n,\mathbb R)$ has determinant one; 
however, this fact does not have a simple proof\upcite{Note:DetSimp}. 
Although (\ref{Eq:SympCond}) seems related to 
${\bf O}^\top {\bf I}_{n} {\bf O} = {\bf I}_{n}$, 
symplectic matrices are in general not isometries since  
$\mathsf S^\top\mathsf S \ne \mathbf I_{2n}$. 
However, a symplectic isometry does exist for the particular case where 
symplectic matrices are also orthogonal\upcite{gossonbook2006,littlejohn1986}.  
It is important to keep in mind that the symplectic group in this work
is defined only for even-dimensional real matrices; that is,   
matrices in ${\rm M}(2n,\mathbb R)$.

In this paper all symplectic matrices, excepting the identity ${\bf I}_{2n}$, 
will be typed with sans-serif fonts, {\it e.g.}, 
$\mathsf{J, S, Z, O, L}$, etc, while all the other matrices appear as 
Roman bold. 

For each real square positive-definite symmetric matrix with {\it even dimension}, 
there is an associated {\it symplectic matrix}  
that diagonalizes it through a {\it congruence relation} in a very specific way. 
This is the content of the Williamson theorem\upcite{simon1999,gossonbook2006,simon1994}: 

\vspace{0.2cm}

\noindent {\bf Theorem:} 
{\it 
Let ${\bf M} \in {\rm M}(2n,\mathbb R)$ be symmetric 
and positive-definite, 
{\it i.e.}, ${\bf M}^\top = {\bf M} > 0$.  
There exists ${\sf S}_{\bf M} \in {\rm Sp}(2n,\mathbb R)$ 
such that 
\begin{equation}                                                      \label{Eq:tw1}      
\begin{aligned}
&{\sf S}_{\bf M} {\bf M} {\sf S}_{\bf M}^\top 
= {\bf \Lambda}_{\bf M}, \\
&{\bf \Lambda}_{\bf M} := 
{\rm Diag}(\mu_1,...,\mu_n,\mu_1,...,\mu_n)  
\end{aligned}
\end{equation}
with
$ 0 < \mu_j \le \mu_k  \,\,\, \text{for} \,\,\, j \le k$. %
Each $\mu_j$ is such that    
\begin{equation}                                                      \label{Eq:tw2}
\det({\sf J} {\bf M} \pm i \mu_j {\bf I}_{2n}) = 0 
\,\,\,\,\,\, (j = 1,...,n), 
\end{equation}
and the matrix ${\sf S}_{\bf M}$ admits the decomposition
\begin{equation}                                                      \label{Eq:tw3}
{\sf S}_{\bf M} = 
\sqrt{{\bf \Lambda}_{\bf M}} \, {\bf O} \, \sqrt{\mathbf{M}^{-1}}, 
\end{equation}
where ${\bf O} \in {\rm M}(2n,\mathbb R)$ satisfies 
\begin{equation}                                                      \label{Eq:tw4}
{\bf O} \, \sqrt{\bf M} \, \mathsf J \, 
\sqrt{\bf M} \, {\bf O}^\top = 
{\bf \Lambda}_{\bf M} \mathsf J , \,\,\,  
\end{equation}
and ${\bf O}^\top = {\bf O}^{-1}$, {\it i.e.}, 
is an orthogonal matrix. }                          
\vspace{0.2cm}

\noindent Before going into the proof, some comments are in order: 

\noindent {---} The matrix ${\sf S}_{\bf M}$ performs a symplectic 
diagonalization through a congruence relation between $\bf M$ and 
${\bf {\bf \Lambda}}_{\bf M}$, 
although generic congruences are not similarity relations. 

\noindent {---} The double-paired ordered set (or the diagonal matrix) 
${\bf \Lambda}_\mathbf{M} \in {\rm M}(2n, \mathbb R)$ 
is called {\it symplectic spectrum} of $\mathbf M$ 
and $\mu_k$ are said to be its {\it symplectic eigenvalues}, 
which are in general not equal to a Euclidean eigenvalue of $\bf M$. 
If in addition 
${\sf S}_{\bf M}^\top = {\sf S}_{\bf M}^{-1}$; 
that is, ${\sf S}_{\bf M}$ is symplectic and orthogonal, 
the matrix ${\bf M}$ will be orthogonally similar to ${{\bf \Lambda}}_{\bf M}$. 
In this situation the symplectic and Euclidean spectrum coincide.

\noindent {---} The complex numbers $\pm i \mu_j$, 
where $\mu_j > 0, \forall j$,  
are the Euclidean eigenvalues of ${\mathsf J} {\bf M}$. 

\noindent {---} The symplectic congruence 
${\bf M}' := {\mathsf S}^\top {\bf M} {\mathsf S}$ for any 
${\mathsf S} \in {\rm Sp}(2n,\mathbb R)$
is equivalent to the similarity 
${\sf J}  {\bf M}' = {\mathsf S}^{-1} {\sf J}  {\bf M} {\mathsf S}$, 
due to the symplectic condition for $\mathsf S$. 
Explicitly, ${\sf J} {\mathsf S}^{\top} {\bf M} {\mathsf S} = 
{\mathsf S}^{-1} {\sf J}  {\bf M} {\mathsf S}$. 

\noindent {---} The symplectic spectrum is invariant 
under symplectic congruences, which means that  
for any ${\mathsf S}\in{\rm Sp}(2n,\mathbb R)$, 
the symplectic spectrum ${\bf \Lambda}_{\mathbf{M}'}$ of 
${\bf M}' := {\mathsf S}^\top {\bf M} {\mathsf S}$ is also 
${\bf \Lambda}_\mathbf{M}$ owing to the similarity 
${\sf J}  {\bf M}' = {\mathsf S}^{-1} {\sf J}  {\bf M} {\mathsf S}$. 


\noindent {---} Due to $\det{\mathsf S}_\mathbf{M} = \det{\mathsf J} =1$, 
then $\det {\bf M} = \det {\bf \Lambda}_\mathbf{M} = 
\det{\sf J}{\bf M} = \mu_1^2\mu_2^2...\mu_n^2$.  
If $n = 1$, ${\bf \Lambda}_\mathbf{M} = \mu_1 {\mathbf I}_2$ and 
$\det {\bf M} = \mu_1^2$.

\noindent {---} The matrix Eq.~(\ref{Eq:tw3}) readily satisfies 
${\sf S}_{\bf M} {\bf M} {\sf S}_{\bf M}^\top = {\bf \Lambda}_{\bf M}$ 
for any orthogonal matrix $\bf O$; 
however ${\sf S}_{\bf M}$ in Eq.~(\ref{Eq:tw3}) 
will be symplectic if and only if the orthogonal 
matrix obeys Eq.~(\ref{Eq:tw4}). 

\noindent {---} There are several situations in physics 
where only the symplectic spectrum of 
a positive-definite matrix $\bf M$ 
is required; following Eq.~(\ref{Eq:tw1}), 
this spectrum is directly obtained through the solution of 
$\det({\mathsf J} {\mathbf M} - \lambda {\mathbf I}_{2n}) = 0$, 
{\it i.e.}, from the {\it Euclidean} eigenvalues 
of ${\mathsf J} {\mathbf M}$.  

\noindent {---} The matrix ${\sf S}_{\bf M}$ 
can be constructed after the determination 
of the symplectic spectrum.  
To this end, the matrix $\bf M$ must be {\it Euclideanly diagonalized} 
and its square root determined.  
To obtain the orthogonal matrix $\bf O$, 
the system of equations in Eq.~(\ref{Eq:tw4}), which has a unique solution for $\bf O$, 
must be solved, and thus Eq.~(\ref{Eq:tw3}) provides the desired symplectic matrix.   

\noindent {---} Squaring both sides of Eq.~(\ref{Eq:tw4}) results in 
$
-{\bf \Lambda}_{\bf M}^2 = {\bf O} (\sqrt{\bf M} \,\mathsf J \,{\bf M}\, \mathsf J \,\sqrt{\bf M})
{\bf O}^\top $, 
which shows that the symmetric matrix in the parentheses is Euclideanly 
diagonalized by the matrix ${\bf O}$. 
The solution of the above eigensystem is in general more convenient 
than solving Eq.~(\ref{Eq:tw4}). 

\vspace{0.2cm}

A detailed proof of the theorem, 
thought to be pedagogical and self-contained, 
is placed in the Supplementary Material\upcite{SupMatDem};
nevertheless an outline (based on 
Ref.\onlinecite{simon1999}) 
may be valuable at this stage. 

\vspace{0.2cm}

\noindent {\bf Outline of the Proof:}
\noindent The main point relies upon the Euclidean diagonalization 
of skewsymmetric matrices, 
in particular the corollary for an even-dimensional 
nonsingular skewsymmetric matrix\upcite{Note:Cor}:   
the matrix $\tilde {\bf M} \in {\rm M}(2n,\mathbb R)$ 
is invertible and skewsymmetric, 
$\tilde {\bf M}^\top = - \tilde {\bf M}$, 
if and only if there is an orthogonal matrix 
${\bf Q} \in {\rm M}(2n,\mathbb R)$  
such that ${\bf Q} \tilde {\bf M} {\bf Q}^\top = 
\mathsf J ({\bf \Omega} \oplus {\bf \Omega})$, 
where $\mathsf J$ is defined in Eq.~(\ref{Eq:SympCond}) 
and ${\bf\Omega} = {\rm Diag}(\omega_1,...,\omega_n)$ with
$\omega_j >0, \forall j$. 

The eigenvalues of $\tilde {\bf M}$ are the roots of 
$\det(\tilde {\bf M} - \lambda {\bf I}_{2n}) = 
\det[\mathsf J ({\bf \Omega} \oplus {\bf \Omega}) 
- \lambda {\bf I}_{2n}] = 0$; 
this last determinant may be evaluated through blocks, 
{\it i.e.}, 
$\det[\mathsf J ({\bf \Omega} \oplus {\bf \Omega}) 
- \lambda {\bf I}_{2n}] = 
\det\left(\begin{smallmatrix}
                -\lambda {\bf I}_n & {\bf \Omega}\\
               -{\bf \Omega} & -\lambda {\bf I}_n
\end{smallmatrix}\right) = 
\det[{\bf \Omega}^2 + \lambda^2 {\bf I}_{n}] = 0$, 
and thus 
the eigenvalues are $\pm i \omega_j$ for $j = 1,...,n$. 

The matrix in Eq.~(\ref{Eq:tw3}) is the most generic matrix 
satisfying Eq.~(\ref{Eq:tw1}), since ${\bf O}$ is a generic orthogonal matrix; 
writing $\tilde {\bf M} := \sqrt{\bf M} \, \mathsf J \, \sqrt{\bf M}$, then 
$\tilde {\bf M} = - \tilde {\bf M}^\top$, since ${\bf M} = {\bf M}^\top$ 
and $\mathsf J^\top = - \mathsf J$. 
Consequently, 
$\det(\tilde {\bf M} - \lambda {\bf I}_{2n}) = 
 \det({\sf J} {\bf M} - \lambda {\bf I}_{2n})$ and the eigenvalues of 
${\sf J} {\bf M}$ will be as the ones above, which is expressed as 
Eq.~(\ref{Eq:tw2}). 

Noting that $\det\tilde{\bf M} = \det {\bf M} > 0$, 
the above corollary is employed and 
it is possible to identify 
$({\bf \Omega} \oplus {\bf \Omega}) = {\bf \Lambda}_{\bf M}$
and ${\bf Q} = {\bf O}$, thus   
${\bf Q} \tilde {\bf M} {\bf Q}^\top = \mathsf J ({\bf \Omega} \oplus {\bf \Omega})$ 
becomes exactly Eq.~(\ref{Eq:tw4}).
Manipulating this last equation, one finds
$\mathsf J = ({\bf O} \sqrt{\bf M} )^{-1} (\sqrt{{\bf \Lambda}_{\bf M}} \,  
\mathsf J \, \sqrt{{\bf \Lambda}_{\bf M}})(\sqrt{\bf M} {\bf O}^\top)^{-1}$, 
which is a symplectic condition for the matrix 
$\mathsf S_{\bf M}$ in Eq.~(\ref{Eq:tw3}). 
\hfill\(\Box\)

\vspace{0.2cm}

In the Supplementary Material\upcite{SupMatDem}
the proof is more detailed and does not assume 
{\it a priori} knowledge of the diagonalization properties 
of skewsymmetric matrices.


\section{Hamiltonian Dynamics}  \label{Sec:HM}                             
The movement of a system in phase space is governed by the Hamilton equations 
\begin{equation}\label{Eq:HamEq}
\dot x = \frac{dx}{dt} = {\mathsf J} \frac{\partial h}{\partial x}, 
\end{equation}                                                                           
where 
$x := (q_1,...,q_n,p_1,...,p_n)^\top$ 
is the vector containing the generalized coordinates and 
momenta of the system, $\mathsf J$ is the symplectic matrix in Eq.~(\ref{Eq:SympCond}), 
and $h = H(x,t)$ is the Hamiltonian of the system.  

A change of variables $x' = f(x,t)$ is said to be {\it canonical} 
if it preserves the equations of motion. 
This will happen if and only if the Jacobian matrix of 
the transformation $\partial f /\partial x \in {\rm M}(2n,\mathbb R)$ 
is a symplectic matrix\upcite{arnold,landau1,goldstein,lemos}; 
that is, if it satisfies Eq.~(\ref{Eq:SympCond}). 
A canonical transformation is linear when the function $f$ is itself 
the linear function $f(x,t) = \mathsf S x$, 
for any symplectic $\mathsf S$.  

As an example\upcite{arnold,landau1,goldstein,lemos,gosson2}  
the one-degree-of-freedom polar (action-angle) transformation 
\begin{equation}
f(x) = (\sqrt{2 q} \cos p,\sqrt{2 q} \sin p)
\end{equation}
is canonical since the Jacobian matrix 
\begin{equation}
\frac{\partial f}{\partial x} =  
\begin{pmatrix}
          \frac{1}{\sqrt{2q}} \cos p   & \frac{1}{\sqrt{2q}} \sin p \\
        - \sqrt{2q} \, \sin p & \sqrt{2q} \, \cos p 
          \end{pmatrix}  
\end{equation}          
is symplectic thanks to $|{\partial f}/{\partial x}| = 1$;  
however $f$ is {\it not} linear. 

For the remainder of this paper, only {\it affine canonical transformations} 
will be relevant. These are compositions of symplectic transformations with  
rigid translations: 
\begin{equation} \label{Eq:affineCT}
f(x,t) = {\mathsf S} x + \eta, 
\end{equation}
for a symplectic $\mathsf S$ and a
$\eta \in {\mathbb R}^{2n}$; 
note that ${\partial f}/{\partial x} = \mathsf S$. 

%

The {\it Poisson bracket} between two functions $f(x,t)$ and $g(x,t)$ 
is written as\upcite{arnold,arnold,landau1,goldstein,lemos}  
\begin{equation}  
\{ f, g \} := 
\mathsf J \frac{\partial f}{\partial x} \cdot \frac{\partial g}{\partial x} \, ;  
\end{equation} 
the presence of the matrix $\mathsf J$ indicates, 
and it is not difficult to show, 
that this structure is invariant under canonical 
transformations\upcite{arnold,arnold,landau1,goldstein,lemos}.  
Choosing $f(x) = x_j$ and $g(x) = x_k$, 
the {\it fundamental Poisson bracket} is obtained 
\begin{equation}\label{Eq:FundPbrack}
\{ x_j, x_k \} = \mathsf J_{jk}. 
\end{equation}
%
%
It is instructive and useful for the next section 
to show the invariance of Eq.~(\ref{Eq:FundPbrack}) under affine 
symplectic transformations.   %
Defining 
$x' := {\mathsf S} x + \eta $,  
then  
\begin{eqnarray}\label{Eq:FundPbrackInv}
\{ x'_j, x'_k \} &=&  \!
\sum_{l,m = 1}^{2n} \! {\mathsf S}_{jl}{\mathsf S}_{km}\{ x_l, x_m \} \nonumber \\ 
&=& ({\mathsf S}\mathsf J{\mathsf S}^\top)_{jk} = \mathsf J_{jk}, 
\end{eqnarray}
due to the symplectic nature of $\mathsf S$. 

\subsection{Quadratic Hamiltonians and Williamson Theorem}\label{Sec:QHWT}   
Consider the time-independent quadratic Hamiltonian 
\begin{equation}\label{Eq:ClQuadLinHam}
H(x) = \tfrac{1}{2} x \cdot \mathbf H x + 
                      x \cdot \xi  + H_0,
\end{equation}
where $\xi \in \mathbb R^{2n}$ is a vector,  
$H_0 \in \mathbb R$ is a constant, and 
$\mathbf H = \mathbf H^\top = \frac{\partial^2 H}{\partial x \partial x}$ 
is the Hessian matrix.  
The corresponding equations of motion follow immediately 
from Eq.~(\ref{Eq:HamEq}) and using that 
$\partial H/\partial x = {\bf H}x + \xi$, yielding 
\begin{equation}\label{Eq:EqMotQuadHam}
\dot x = {\mathsf J} \mathbf H \, x + {\mathsf J} \xi.     
\end{equation} 
If $\det {\bf H} \ne 0$, 
a direct substitution shows that the solution of Eq.~(\ref{Eq:EqMotQuadHam}) is given by
\begin{equation}\label{Eq:FluxQuadLinHam}
x(t) = {\mathsf S}_t (x_0 + {\mathbf H}^{-1} \xi) 
- {\mathbf H}^{-1}\xi, 
\,\,\,\,\,
\mathsf S_t := {\exp}[{\mathsf J {\bf H} t}] 
\end{equation} 
for an initial condition $x_0 := x(0)$. 
The phase-space point $x_\star := - {\mathbf H}^{-1}\xi$ 
is an equilibrium (or fixed) point of the system, 
since $x(t) = x_\star, \forall t$, 
if $x_0 = x_\star$.   
Even when $\det {\bf H} = 0$, 
an analytical solution like Eq.~(\ref{Eq:FluxQuadLinHam}) 
can be obtained; see the Supplementary Material\upcite{SupMatQH}.  

Due to the symmetricity of $\bf H$, 
the above defined matrix $\mathsf S_t$ 
is itself a symplectic matrix, since 
\begin{eqnarray}\label{Eq:SimpEv}
{\mathsf J} {\mathsf S}_t {\mathsf J}^{-1} &=&
{\exp}[ {\mathsf J}^2 {\bf H} {\mathsf J}^{-1} t] =
{\exp}[ - ({\mathsf J}{\bf H})^{\top} t]   \nonumber \\  
&=&  
( {\exp}[ - {\mathsf J}{\bf H} t] )^\top = 
({\mathsf S}_t^{-1})^\top = ({\mathsf S}_t^\top)^{-1},  
\end{eqnarray}
where we used that 
$\mathsf J^\top = \mathsf J^{-1} = - \mathsf J$ and 
${\exp}({\bf A}\!^\top) = (\exp{\bf A})^\top$; 
multiplying the above equation by ${\mathsf S}_t^\top$ from the left and by 
$\mathsf J$ from the right, the symplectic condition in Eq.~(\ref{Eq:SympCond}) is obtained. 
Note also that ${\mathsf S}_{-t} = {\mathsf S}_t^{-1}$. 
It is noteworthy that 
the temporal evolution in Eq.~(\ref{Eq:FluxQuadLinHam}) 
is an affine canonical transformation, 
as defined in Eq.~(\ref{Eq:affineCT}). 
All these properties remain valid for any matrix $\bf H$; 
see the Supplementary Material\upcite{SupMatQH}.  

Regardless of the analytic solution for a generic quadratic Hamiltonian, 
the behavior of the system (or the matrix $\mathsf S_t$) 
can be very awkward due to the exponential structure in Eq.~(\ref{Eq:FluxQuadLinHam}), 
even considering $\det {\bf H} \ne 0$. 
Fortunately, the Williamson theorem is useful to simplify 
the description of the system's behavior when ${\bf H}$ is positive-definite.   

Considering ${\bf H} > 0$,  Eq.~(\ref{Eq:tw1}) can be applied, 
\begin{equation}\label{Eq:WTHam}
{\sf S}_{\bf H} {\bf H} {\sf S}_{\bf H}^\top = {\bf \Lambda}_{\bf H},
\end{equation} 
and the Hamiltonian Eq.~(\ref{Eq:ClQuadLinHam}) becomes 
\begin{eqnarray}
H(x) &=& \tfrac{1}{2} x \cdot {\sf S}_{\bf H}^{-1} 
                              {\bf \Lambda}_{\bf H} {\sf S}_{\bf H}^{-\top} x
                            + x \cdot \xi  + H_0     \nonumber \\
     &=& \tfrac{1}{2} {\sf S}_{\bf H}^{-\top} x \cdot 
                      {\bf \Lambda}_{\bf H} {\sf S}_{\bf H}^{-\top} x
                     + {\sf S}_{\bf H}^{-\top} x \cdot {\sf S}_{\bf H} \xi  
                     + H_0,  
\end{eqnarray}
where, for compactness, we introduced the notation 
${\bf A}^{-\top} := ({\bf A}^{\!\top})^{-1} = 
({\bf A}^{-1})^{\top}$. 
Through the theorem, 
any quadratic Hamiltonian with a positive-definite Hessian describes a collection of 
$n$ harmonic oscillators, 
since performing the canonical transformation 
$x' = {\sf S}_{\bf H}^{-\top}\, x$, for ${\sf S}_{\bf H}$ in Eq.~(\ref{Eq:tw3}),  
the Hamiltonian of the system turns into
\begin{eqnarray} \label{Eq:NmHam}
H'(x') &:=& H({\sf S}_{\bf H}^\top x') = \tfrac{1}{2} x' \cdot {\bf \Lambda}_{\bf H} x' 
      + x' \cdot {\sf S}_{\bf H} \xi  + H_0   \nonumber  \\
&=& \tfrac{1}{2} (x'-x'_\star) \cdot {\bf \Lambda}_{\bf H} (x'-x'_\star) + H'_0,
\end{eqnarray}
where 
$H'_0 := H_0 - \tfrac{1}{2} x'_\star \cdot {\bf \Lambda}_{\bf H} x'_\star$ 
is an (constant) offset of the Hamiltonian,  
$x'_\star := {\sf S}_{\bf H}^{-\top} x_\star = 
- {\bf \Lambda}_{\bf H}^{-1} {\sf S}_{\bf H} \xi$ 
is the equilibrium coordinate (fixed point of $H'$), 
and, from Eq.~(\ref{Eq:tw1}), the quadratic form is 
\begin{eqnarray}
\tfrac{1}{2} (x'-x'_\star) \cdot {\bf \Lambda}_{\bf H} (x'-x'_\star) = &&\nonumber\\ 
\sum_{k = 1}^n \frac{\mu_k}{2}({p}_k' - {p}_{\star k}' )^2 &&
             + \frac{\mu_k}{2}({q}_k' - {q}_{\star k}' )^2. 
\end{eqnarray}

The most important consequence of the Williamson theorem 
is expressed in the linear canonical transformation 
$x' = {\sf S}_{\bf H}^{-\top}\, x$, 
which brings the system to its normal-mode phase-space coordinates 
and reveals the eigenfrequencies of the system to be the symplectic eigenvalues
contained in ${\bf \Lambda}_{\bf H}$.
Writing the equations of motion for the normal modes, 
{\it i.e.}, performing the transformation $x' = {\sf S}_{\bf H}^{-\top}\, x$
in Eq.~(\ref{Eq:HamEq}), the Hamilton equation becomes 
\begin{equation}\label{Eq:HamEqNm}
\dot x' = {\mathsf J} \,\frac{\partial h'}{\partial x'} = 
{\mathsf J} {\bf \Lambda}_{\bf H} (x'-x'_\star), 
\end{equation} 
for $h' = H'(x')$ in Eq.~(\ref{Eq:NmHam}), 
with solution given by  
\begin{equation}\label{Eq:NmFlux}
x'(t) = {\mathsf S}'_t (x'_0 - x'_\star) + x'_\star, \,\,\, 
{\mathsf S}'_t := {\exp}[{\mathsf J {\bf \Lambda}_{\bf H} t}].
\end{equation} 
Recalling that ${\mathsf J}^2 = -{\bf I}_{2n}$, 
the evolution matrix can be cast into the form
\begin{equation}\label{Eq:NmEvol}
{\mathsf S}'_t = {\exp}[{\mathsf J {\bf \Lambda}_{\bf H} t}] = 
\cos({\bf \Lambda}_{\bf H} t) + {\mathsf J} \sin({\bf \Lambda}_{\bf H} t),  
\end{equation}
since 
\begin{eqnarray}
&&{\exp}[{\mathsf J {\bf \Lambda}_{\bf H} t}] = 
\sum_{k=0}^{\infty} 
\left[ \frac{({\mathsf J {\bf \Lambda}_{\bf H} t})^{2k}}{(2k)!} + 
        \frac{({\mathsf J {\bf \Lambda}_{\bf H} t})^{2k+1}}{(2k+1)!} \right] \nonumber \\
&&=
\sum_{k=0}^{\infty} 
\left[ (-1)^k \frac{({{\bf \Lambda}_{\bf H} t})^{2k}}{(2k)!} + 
(-1)^{k} \mathsf J \frac{({ {\bf \Lambda}_{\bf H} t})^{2k+1}}{(2k+1)!} \right].
\end{eqnarray}
The symplectic matrix ${\mathsf S}'_t$ is also orthogonal,
\begin{equation}
{{\mathsf S}'_t}^{-1} = {{\mathsf S}'_{-t}} =  
{\exp}[-{\mathsf J {\bf \Lambda}_{\bf H} t}] =
{\exp}[{\bf \Lambda}_{\bf H} {\mathsf J^\top  t}] = {{\mathsf S}'_{t}}^\top, 
\end{equation}
and the evolution of the system in Eq.~(\ref{Eq:NmFlux}) is thus a rotation in 
phase space around the equilibrium point $x'_\star$, 
where each conjugate pair evolves as 
\begin{equation}\label{Eq:NmCompEv}
\left( \!\!\begin{array}{c} 
          q'_k(t) - q'_{\star k} \\
          p'_k(t) - p'_{\star k}
       \end{array} \!\!\right) \!= \!
\left(\!\! \begin{array} {rc} 
          \cos \mu_k t & \sin \mu_k t  \\
          -\sin \mu_k t & \cos \mu_k t
\end{array}\!\! \right) \!\!
\left(\!\! \begin{array}{c} 
          q'_{0k} - q'_{\star k} \\
          p'_{0k} - p'_{\star k}
\end{array}\!\! \right). 
\end{equation}

The solution of the original system is recovered performing the 
inverse transformation $x = {\sf S}_{\bf H}^{\top}\, x'$, giving 
$x(t) = {\sf S}_{\bf H}^{\top}\, x'(t)$ for $x'(t)$ in Eq.~(\ref{Eq:NmFlux}), 
which is precisely Eq.~(\ref{Eq:FluxQuadLinHam}) since 
\begin{eqnarray}\label{Eq:SympNmTrans}
{\mathsf S}_t &=& {\exp}[{\mathsf J {\bf H} t}] = 
{\exp}[{\mathsf J {\sf S}_{\bf H}^{-1} 
       {\bf \Lambda}_{\bf H} {\sf S}_{\bf H}^{-\top}t} ] \nonumber \\
&=&{\exp}[{{\sf S}_{\bf H}^{\top}\mathsf J 
           {\bf \Lambda}_{\bf H} {\sf S}_{\bf H}^{-\top}t}] 
= {\sf S}_{\bf H}^{\top} \,
{\exp}[{\mathsf J  {\bf \Lambda}_{\bf H} t}] 
{\sf S}_{\bf H}^{-\top} \nonumber \\
&=& {\sf S}_{\bf H}^{\top} {\mathsf S}'_t {\sf S}_{\bf H}^{-\top} .
\end{eqnarray}

As a last comment, 
the Hamiltonian in Eq.~(\ref{Eq:ClQuadLinHam}) can be conveniently rewritten as 
\begin{equation}
H(x) = \tfrac{1}{2} (x - x_\star) \cdot \mathbf H (x -  x_\star) 
- \tfrac{1}{2} \xi \cdot {\mathbf H}^{-1} \xi + H_0
\end{equation}
%
and the affine transformation 
\begin{equation}\label{Eq:affineTrans}
x''= {\sf S}_{\bf H}^{-\top}( x - x_\star)
\end{equation}
reduces the above Hamiltonian to 
$H''(x'') = \tfrac{1}{2} x''\cdot {\bf \Lambda}_{\mathbf H} x'' 
+ H_0'$,   
which describes oscillations as in Eq.~(\ref{Eq:NmHam}),  
but around the origin of phase space. 
The reason to keep the equilibrium coordinate $x'_\star$ in Eq.~(\ref{Eq:NmHam}) 
is related to the study of small oscillations,  
where the Hamiltonian often has multiple fixed points 
and it may be interesting to analyze the behavior of the system around 
each of them, as will become clear soon.  
Nevertheless, $H''(x'')$ can always be obtained performing 
the (canonical) rigid translation 
$ (x'-x'_\star) \mapsto x''$ in Eq.~(\ref{Eq:NmHam}).  
%

\subsection{Complex Phase-Space} 
The resemblance of Eq.~(\ref{Eq:NmEvol}) to the Euler formula, 
${\rm e}^{i\theta} = \cos\theta + i \sin\theta$, is noticeable. 
In the former, 
the matrix $\mathsf J$ is such that $\mathsf J^2 = -{\bf I}_{2n}$ 
and performs the role of the imaginary unity. 
The mechanical Euler-like behavior can be further explored by diagonalizing 
the matrix $\mathsf J {\bf \Lambda}_{\bf H}$: 
\begin{equation}\label{Eq:CompDiag}
  {\mathbf W} (\mathsf J {\bf \Lambda}_{\bf H})   {\mathbf W}^\dag = i 
\left(\begin{array}{cc}
{\bf \Omega} & {\bf 0}_n  \\
{\bf 0}_n & - {\bf \Omega} 
\end{array}\right), 
\end{equation}
where ${\bf \Omega} := {\rm Diag}(\mu_1,...,\mu_n)$ and  the unitary matrix 
\begin{equation}\label{Eq:CompSymp}
  {\mathbf W} := \frac{1}{\sqrt{2}}
\left(\begin{array}{cc}
       {\mathbf I}_{n}   & i {\mathbf I}_{n} \\
       i {\mathbf I}_{n} & {\mathbf I}_{n}      
      \end{array} \right)
%
\end{equation}
is symmetric ${\mathbf W}^\top = {\mathbf W}$. Note that 
${\mathbf W}^\dagger = {\mathbf W}^\ast = {\mathbf W}^{-1}$ and  
$ {\mathbf W}^\top \!\mathsf J {\mathbf W} = \mathsf J$. 
Last property is the condition (\ref{Eq:SympCond}) for the complex matrix $\bf W$,  
however the symplectic group is only defined for real matrices. 

Considering the vectors $q'= (q'_1,...,q'_n)^\top$ and 
$p' = (p'_1,...,p'_n)^\top $,  
the canonical complex change of coordinates 
\begin{equation} \label{Eq:CompTrans}
z :=   {\mathbf W} x' = \frac{1}{\sqrt{2}}\left( \begin{array}{c} 
                                              q' + i p'         \\                          
                                              iq' -  p'
                                              \end{array} \right) 
\end{equation}
transforms the equations of motion Eq.~(\ref{Eq:NmFlux}) to 
\begin{equation} \label{Eq:CompSol}
\begin{aligned}
z(t) &= \tilde{\mathbf S}_t (z_0 - z_\star) + z_\star, \\
\tilde{\mathbf S}_t &:=   {\mathbf W} {\mathsf S}'_t   {\mathbf W}^\ast = 
{\rm e}^{i {\bf \Omega}t} \oplus {\rm e}^{-i {\bf \Omega}t},   
\end{aligned}
\end{equation} 
where we used Eq.~(\ref{Eq:CompDiag}). 
Each component ($ k = 1,...,n$) in the previous equation evolves as
$(z_k(t) - {z_\star}_k) = {\rm e}^{i \mu_k t} ({z_0}_k - {z_\star}_k)$, 
%
which is the complex version of Eq.~(\ref{Eq:NmCompEv}). 
Despite complex, since ${\bf W} \in {\rm Mat}(2n,\mathbb C)$, 
transformation Eq.~(\ref{Eq:CompTrans}) preserves not only the Poisson bracket, 
as in Eq.~(\ref{Eq:FundPbrackInv}), but also the Hamilton's equations, 
$\dot{z} = {\mathsf J} \partial \tilde h/\partial z$, where 
$\tilde h = H'(  {\mathbf W}^\ast x')$ 
with $H'$ given by Eq.~(\ref{Eq:NmHam}).
As we shall see, transformation Eq.~(\ref{Eq:CompTrans}) is the bridge towards 
the creation-annihilation operators in quantum mechanics 
and the coordinates $z$ are their classical counterpart.  

\subsection{The Problem of Small Oscillations}\label{Sec:TPSO}             
Consider a generic time-independent Hamiltonian $h$ described 
by a smooth function $H(x)$. 
A fixed point of the system, denoted $x_\star$, 
is an initial condition that does not evolve: 
$x(t) = x_\star, \forall t \in \mathbb R$, 
which can be determined by the solution of 
\begin{equation}
\dot x = 0 \Longleftrightarrow 
{\mathsf J} \frac{\partial h}{\partial x}\Big|_{x = x_\star} = 0 .    
\end{equation}
The behavior of the system around the fixed point can be determined by 
a Taylor expansion up to second order:
\begin{equation}\label{Eq:HamExp}
H(x) \approx H(x_\star) + \xi_\star \cdot (x-x_\star) + 
\tfrac{1}{2} (x-x_\star)\cdot {\bf H}_\star (x-x_\star),    
\end{equation}
where
\begin{equation}
\begin{aligned}
\xi_\star &:= \frac{\partial H}{\partial x}\Big|_{x=x_\star} \in {\mathbb R}^n, \\
{\bf H}_\star &:= \frac{\partial^2 H} {\partial x \partial x} \Big|_{x=x_\star}  
               \in {\rm M}(2n,\mathbb R). 
\end{aligned}
\end{equation}
This approximation leads to a 
quadratic Hamiltonian like Eq.~(\ref{Eq:ClQuadLinHam}) 
and the solution around the fixed point 
is determined by Eq.~(\ref{Eq:FluxQuadLinHam}).
If ${\bf H}_\star > 0$, the movement of the system is described 
by the analysis already performed with the Williamson theorem. 

In principle, 
the problem of small oscillations is solved as described in Sec.\ref{Sec:QHWT}. 
However, it is worth emphasizing that the efficiency of 
the approximation Eq.~(\ref{Eq:HamExp}) 
is only guaranteed if the trajectories of the original system 
always remain close to $x_\star$, 
which is equivalent to saying that 
the fixed point is a {\it stable center}\upcite{arnoldODE}. 
For a 
quadratic Hamiltonian of the form Eq.~(\ref{Eq:ClQuadLinHam}), 
a necessary and sufficient condition for this stability is ${\bf H}>0$.    
However, considering generic Hamiltonians, 
there are situations where the stability will depend on higher-order terms, 
which includes the case in which ${\bf H}_\star = 0$, 
and the present theory does not apply\upcite{Note:Anom}.
%
%
The analysis for generic systems 
is a subject of the Lyapunov stability theory\upcite{arnoldODE} 
and is far from the objectives of this paper.  

Other kinds of expansions can be performed on a generic 
Hamiltonian and the Williamson theorem can be also 
useful to describe the behavior of the system. 
For instance, if the Hamiltonian depends on a parameter $\epsilon$, 
an expansion like 
\begin{equation}
H(x,\epsilon) = \sum_{k = 0}^{\infty} 
\frac{\epsilon^k}{k!}
\frac{\partial^k\!H}{\partial\epsilon^k}\bigg|_{\epsilon = 0}  
\end{equation}
will be structurally different from Eq.~(\ref{Eq:HamExp}), 
although it can also provide a quadratic Hamiltonian if truncated\upcite{SupMatEx2}.
The above stability discussion can be translated 
to the present case if the fixed points of the truncated expansion remain close 
to the ones of the original Hamiltonian. 

In the Supplementary Material\upcite{SupMatLag} the Lagragian way of dealing 
with oscillations\upcite{arnold,landau1,goldstein,lemos} is straightforwardly 
developed and compared with the Hamiltonian description. 
The advantages of the latter becomes clear since the Williansom theorem 
enables the treatment of more general systems. 

\section{Quantum Mechanics} \label{Sec:QM}
In quantum mechanics, classical observables (functions of position and momenta) 
are promoted to operators, or linear functions, acting on the Hilbert space of the 
quantum system $\mathcal H$. 
A system of $n$ degrees of freedom is thus described by position and momenta operators, 
which will be collectively written as operator vectors\upcite{Note:VecOp}: 
\begin{equation}\label{Eq:xop}
\hat x := \left( \begin{array}{c}
                 \hat q_1 \\ \vdots \\ \hat q_n  \\ 
                 \hat p_1 \\ \vdots \\ \hat p_n 
                 \end{array}\right), \,\,\,
\hat x^\top := \left( \hat q_1, ..., \hat q_n, \hat p_1, ..., \hat p_n \right).
\end{equation}
Note that the action of ``$\top$'' on the operator vector means
the usual vector transposition. 
The ``scalar" (dot) product between two of these vectors is 
\begin{equation}
\hat x \cdot \hat y := \hat x^\top \hat y = \sum_{j = 1}^{2n} \hat x_j\hat y_j. 
\end{equation}

The canonical commutation relation $[\hat q_l, \hat p_m] = i \hbar \delta_{lm}$
is translated to the collective notation as 
\begin{equation} \label{Eq:CCR}
[\hat x_j, \hat x_k] = 
i \hbar \mathsf J_{jk} \,\,\,\,\, (j,k = 1,...,2n) .
\end{equation}
The great advantage of this notation is apparent:  
Like the fundamental Poisson bracket Eq.~(\ref{Eq:FundPbrack}),
the above commutator is invariant under affine symplectic transformations. 
Indeed, defining the new operator vector as 
$\hat x' = {\mathsf S} \hat x + \eta$,  
one obtains\upcite{Note:VeOp2} just as in Eq.~(\ref{Eq:FundPbrackInv}) that
\begin{equation}\label{Eq:InvCCR}
\![ \hat x'_j, \hat x'_k ] =  \!\!\!
\sum_{l,m = 1}^{2n} \!\! 
{\mathsf S}_{jl}{\mathsf S}_{km}[ \hat x_l, \hat x_m ] = \!
i\hbar ({\mathsf S}\mathsf J{\mathsf S}^\top)_{jk} = i\hbar \mathsf J_{jk} \, .
\end{equation}
This invariance highlights that symplectic matrices also 
play a special role in quantum mechanics 
and further one can say that quantum mechanics inherits 
the symplectic structure of classical phase space.   
But how do (affine) symplectic transformations arise in quantum mechanics?
The answer is, as it will be seen, in the same way as in classical dynamics, 
{\it i.e.}, solving equations of motion for a quadratic Hamiltonian. 

Consider a Hamiltonian $\hat h = H(\hat x)$ 
where $H$ is given by Eq.~(\ref{Eq:ClQuadLinHam}). 
The Heisenberg equation of motion\upcite{sakurai,ballentine,cohen} 
for the operator $\hat x$ is 
\begin{equation}\label{Eq:HeisEvol}
\begin{aligned}
\frac{d \hat x_j }{dt} &= \frac{i}{\hbar} 
\left[ H(\hat x), \hat x_j\right]= 
\frac{i}{\hbar}   
\left[ \tfrac{1}{2}\hat x \cdot {\bf H} \hat x + \xi \cdot \hat x + H_0, \hat x_j\right] \\ 
&= \sum_{k = 1}^{2n}
(\mathsf J {\bf H})_{jk} \hat x_k + \sum_{k = 1}^{2n} {\mathsf J}_{jk}\xi_k, 
\end{aligned}
\end{equation}
where $j =1,...,2n$. 
The previous commutator is evaluated using only the
canonical commutation relation Eq.~(\ref{Eq:CCR}). 
Indeed, 
$
\left[\hat x_k \xi_k , \hat x_j\right] =
{\xi}_{k}(\hat x_k \hat x_j - \hat x_j \hat x_k ) = 
i\hbar {\mathsf J}_{kj}{\xi}_{k} 
$
and 
\begin{eqnarray}
\left[\hat x_k {\bf H}_{kl} \hat x_l , \hat x_j\right] &=& 
{\bf H}_{kl}(\hat x_k \hat x_l \hat x_j - \hat x_j \hat x_k \hat x_l ) \nonumber \\ 
&=&
{\bf H}_{kl}(\hat x_k \hat x_j \hat x_l + i\hbar {\mathsf J}_{lj} \hat x_k   - 
            \hat x_j \hat x_k \hat x_l ) \nonumber \\  
&=&
{\bf H}_{kl}(i\hbar {\mathsf J}_{kj} \hat x_l + i\hbar {\mathsf J}_{lj} \hat x_k ) \nonumber \\ 
&=& -i\hbar ({\mathsf J}_{jk}{\bf H}_{kl} \hat x_l + {\mathsf J}_{jl}{\bf H}_{lk} \hat x_k), 
\end{eqnarray} 
where last equality is attained using that 
${\bf H}^\top = {\bf H}$ and ${\sf J}^\top = -{\sf J}$. 

The Heisenberg equation in Eq.~(\ref{Eq:HeisEvol}) is 
exactly the Hamilton equation, Eq.~(\ref{Eq:EqMotQuadHam}), 
with the replacement $x \mapsto \hat x$. 
Thus, from Eq.~(\ref{Eq:FluxQuadLinHam}), its solution is
\begin{equation}\label{Eq:FluxQuantQuadHam}
\hat x(t) = {\mathsf S}_t (\hat x_0 + {\mathbf H}^{-1} \xi) 
- {\mathbf H}^{-1}\xi.  
\end{equation}
The very same treatment is suitable also for the general quadratic case, 
where $\bf H$ may not be positive-definite,
see the Supplementary Material\upcite{SupMatQH}. 
\subsection*{Quantum Normal Modes}                                         
For a positive-definite matrix $\bf H$, 
the Williamson theorem can be applied as in Eq.~(\ref{Eq:WTHam}), and
the solution in Eq.~(\ref{Eq:FluxQuantQuadHam}) can be brought 
to the normal-mode coordinates through the symplectic transformation 
$\hat x' := {\mathsf S}_{\bf H}^{-\top} \hat x$.  
Indeed, 
\begin{eqnarray}\label{Eq:QuaNm}
\hat x'(t) &=& 
{\mathsf S}_{\bf H}^{-\top} {\mathsf S}_t ({\mathsf S}_{\bf H}^{\top}\hat x'_0 
+ {\mathbf H}^{-1} \xi) 
- {\mathsf S}_{\bf H}^{-\top} {\mathbf H}^{-1}\xi \nonumber \\
&=&
{\mathsf S}'_t (\hat x'_0 - x'_\star) + x'_\star, 
\end{eqnarray}
where Eq.~(\ref{Eq:SympNmTrans}) was employed,  
${\mathsf S}'_t$ is written in Eq.~(\ref{Eq:NmEvol}), and 
$x'_\star$ is defined below Eq.~(\ref{Eq:NmHam}).

Thanks to the commutation relation, Eq.~(\ref{Eq:CCR}), 
which is responsible for the coincidence of the Heisenberg equation, Eq.~(\ref{Eq:HeisEvol}), 
with the Hamilton equation, Eq.~(\ref{Eq:EqMotQuadHam}),
all the treatment performed in Sec.~\ref{Sec:QHWT} is precisely the same:  
all equations and results remain valid through the quantization 
$x \mapsto \hat x$. Equations (\ref{Eq:FluxQuantQuadHam}) and (\ref{Eq:QuaNm}) 
are only two examples of this fact. 
For instance, 
the reader is invited to perform the transformation 
$\hat x' := {\mathsf S}_{\bf H}^{-\top} \hat x$ on the 
Heisenberg equation, Eq.~(\ref{Eq:HeisEvol}), 
to obtain the quantum counterpart of Eq.~(\ref{Eq:HamEqNm}). 
This is also true when considering the problem of small oscillations: 
the description in Sec.~\ref{Sec:TPSO} can be rigorously 
translated to the quantum case when replacing the 
Hamiltonian by its quantum version\upcite{Note:QuantProb} 
$\hat h = H(\hat x)$ for a smooth function $H$. 

Quantum oscillators are generally treated in the framework of
creation and annihilation operators\upcite{sakurai,ballentine,cohen}.
For a system of $n$ degrees of freedom, it is convenient to define
a collective notation for these operators through the vector 
\begin{equation} \label{Eq:zdef}
\begin{aligned}
\hat z &:= \sqrt{\hbar} \left(\! \begin{array}{c}
                              \hat a_1 \\ \vdots \\ \hat a_n  \\ 
                              i \hat a_1^\dag \\ \vdots \\ i \hat a_n^\dag 
                              \end{array}\!\right), \\  
\hat z^\dagger &:= \sqrt{\hbar}\,   
           (   \hat a_1^\dagger,..., \hat a_n^\dagger, 
            -i \hat a_1,..., -i \hat a_n ),
\end{aligned}
\end{equation}
where $\hat a_j$ (resp.~$\hat a_j^\dag$) is the creation (resp. annihilation) 
operator of an oscillator with mass $m_j$ and frequency $\omega_j$, namely, 
$\hat a_j := 
\sqrt{\tfrac{m_j \omega_j }{2\hbar}} \, \hat q_j + 
i \sqrt{\tfrac{1}{2\hbar m_j \omega_j}} \, \hat p_j$. 
Observe that the adjoint operation ``$\dag$'' acting on the vector $\hat z$ is twofold: 
it means the ordinary vector transposition together with the 
Hermitian conjugation of each vector component. 
In this way, the ``scalar'' product between two of these vectors, say $\hat z$ and $\hat w$, 
is 
\begin{equation}
\hat z^\dag \hat w := \sum_{j = 1}^n \hat z_k^\dag \hat w_k,  
\end{equation}
and note that, for the operator $\hat x$ in Eq.~(\ref{Eq:xop}), 
$\hat x^\dag = \hat x^\top$. 

The relation between $\hat z$ and $\hat x$ is the complex linear transformation  
\begin{equation}\label{Eq:ComplRep}
\hat z =    {\mathbf W} \, {\mathsf Z} \, \hat x,  
\end{equation}
where $  {\mathbf W}$ is the unitary matrix in Eq.~(\ref{Eq:CompSymp}) 
and ${\mathsf Z}$ is the real symmetric symplectic matrix 
\begin{equation}
{\mathsf Z} := {\rm Diag}\!
\left( \sqrt{\scriptstyle{m_1 \omega_1}}, ..., \sqrt{\scriptstyle{m_n \omega_n}} , 
       \tfrac{1}{\sqrt{m_1 \omega_1}} ,..., \tfrac{1}{\sqrt{m_n \omega_n}} \right). 
\end{equation}
Since ${\mathbf W} {\mathsf J} {\mathbf W} = {\mathsf J}$, 
the same steps in Eq.~(\ref{Eq:InvCCR}) lead from Eq.~(\ref{Eq:CCR}) to 
\begin{equation} \label{Eq:CCRz}
[\hat z_j, \hat z_k] = i \hbar \mathsf J_{jk} \,\,\, (j,k = 1,...,2n),    
\end{equation}
which is equivalent to $[\hat a_j,\hat a_k^\dagger] = \delta_{jk}$,  
and shows that the complex ``coordinates'' $\hat z$ 
constitute a canonical system. 
Note that $\hat z$ has a very particular structure in Eq.~(\ref{Eq:zdef}); 
the factor $\sqrt{\hbar}$ and the imaginary $i$'s explicitly written in this
equation are 
responsible for the canonical structure
of the commutation relation Eq.~(\ref{Eq:CCRz}). 

Matrix ${\mathsf Z}$ 
represents a simultaneous change of units for position and momentum. 
It is useful for the construction of creation-annihilation operators related to 
given oscillators, which are characterized by a given set of masses and frequencies. 
Symplectically equivalent creation-annihilation operators can be constructed  
using $\hat z' =  {\mathbf W} \, {\mathsf S} \hat x$ for any symplectic $\sf S$. 
In particular for 
${\mathsf S} = {\sf S}_{\bf H}^{-\top}$, 
the vector operator $\hat z' = {\mathbf W}{\sf S}_{\bf H}^{-\top} \hat x$ 
is the quantization of Eq.~(\ref{Eq:CompTrans}). 
It is important to stress that transformations Eq.~(\ref{Eq:ComplRep}) and Eq.~(\ref{Eq:CompTrans})
can be applied to any physical system described by coordinates and momenta,
not only the oscillatory ones. 

The quadratic Hamiltonian $\hat h = H(\hat x)$ with $H$ given by Eq.~(\ref{Eq:ClQuadLinHam}) 
through the transformation Eq.~(\ref{Eq:ComplRep}) becomes 
${\tilde h} = H(\mathsf Z^{-1} {\mathbf W}^\ast \hat z)$.  
Noting that 
$\hat x \cdot \eta =\hat x^\top \eta= \hat x^\dagger \eta$ for any real vector $\eta$,  
the new Hamiltonian can be written as 
\begin{equation}\label{Eq:ComQuantQuadHam}
{\tilde h} = 
\tfrac{1}{2}\hat z^\dag \tilde {\bf H} \hat z 
+ \hat z^\dag\zeta + H_0,
\end{equation}
where 
\begin{equation}\label{Eq:CompHess}
\begin{aligned}
\tilde {\bf H} &:= 
( {\mathbf W} {\mathsf Z}^{-1}) \, {\bf H} \, 
( {\mathbf W} {\mathsf Z}^{-1})^\dag = 
\tilde {\bf H}^\dag \in {\rm M}(2n,\mathbb C),  \\ 
\zeta &:=   ({\mathbf W}{\mathsf Z}^{-1})\xi \in {\mathbb C}^{2n}. 
\end{aligned}
\end{equation}
Transformation Eq.~(\ref{Eq:ComplRep}) preserves the Hermitian 
character of the Hamiltonian, 
since $(\hat z^\dag \tilde {\bf H} \hat z)^\dag = 
\hat z^\dag \tilde {\bf H} \hat z $ 
and $(\hat z^\dag \zeta)^\dag = \zeta^\dag \hat z = \hat z^\dag \zeta$ 
for the above defined vector $\zeta$. 

As before, the canonical structure in Eq.~(\ref{Eq:CCRz}) ensures 
that the treatment for quadratic Hamiltonians are readily translated to 
the new set of variables $\hat z$;,
however, now with complex matrices and vectors. 
For instance, the solution Eq.~(\ref{Eq:FluxQuantQuadHam}) 
under the change of variables Eq.~(\ref{Eq:ComplRep}) becomes
\begin{equation}\label{Eq:ComplFlux}
\hat z(t) = 
\tilde {\mathbf S}_t (\hat z_0 - z_\star) + z_\star, 
\end{equation}
where $z_\star := -\tilde {\bf H}^{-1}\zeta$ and  
$
\tilde {\mathbf S}_t :=  ({\mathbf W} {\mathsf Z}) {\mathsf S}_t  
                         ({\mathbf W}{\mathsf Z})^{-1}
                      = {\rm e}^{\mathsf J \tilde{\bf H} t}$,
for ${\mathsf S}_t$ in Eq.~(\ref{Eq:FluxQuadLinHam}). 
Note that 
$\tilde {\mathbf S}_t^\top \mathsf J \tilde {\mathbf S}_t = \mathsf J$.

The Williamson theorem is applicable only to real matrices and,  
once a system is described by a Hamiltonian written as Eq.~(\ref{Eq:ComQuantQuadHam}), 
some adaptations are needed. 
Of course, the inverse of transformation Eq.~(\ref{Eq:ComplRep}) 
can always be applied to Eq.~(\ref{Eq:ComQuantQuadHam}) 
and the transformed Hamiltonian could be analyzed as before.
Nonetheless, a straightforward approach is desirable 
since creation-annihilation operators are ubiquitous in physics. 

The real and complex Hessians in Eq.~(\ref{Eq:CompHess}) are related by a congruence, 
thus $\tilde {\bf H} > 0 \Longleftrightarrow {\bf H} > 0 $. 
For a positive-definite ${\bf H}$, 
Eq.~(\ref{Eq:tw1}) is equivalent to 
\begin{equation}
\tilde{\mathbf S}_{\bf H} 
\tilde{\bf H} \tilde{\mathbf S}_{\bf H}^\dag = 
{\bf W} {\bf \Lambda}_{\bf H}  {\bf W}^\ast = {\bf \Lambda}_{\bf H}, 
\end{equation}
where $\tilde{\mathbf S}_{\bf H} := 
{\bf W} ({\mathsf S}_{\bf H} {\mathsf Z}){\bf W}^\ast$. 
The last diagonalization relation induces the change of variables 
\begin{equation} \label{Eq:BogTrans}
\hat z' := \tilde{\mathbf S}_{\bf H}^{-\dag} \hat z 
\end{equation}
to be implemented in solution Eq.~(\ref{Eq:ComplFlux}).  
Noting that 
$\tilde{\mathbf S}_{\bf H}^{-\dag} 
{\sf J} \tilde{\bf H} \tilde{\mathbf S}_{\bf H}^{-\dag} = 
{\sf J}{\bf \Lambda}_{\bf H}$,   
the mentioned equation reads
\begin{eqnarray} \label{Eq:CompDiagSol}
(\hat z'(t) - z'_\star) &=& 
{\bf W} \exp[\mathsf J {\bf \Lambda}_{\bf H} t] {\bf W}^\ast 
(\hat z_0' - \hat z_\star') \nonumber  \\ 
&=& 
( {\rm e}^{i {\bf \Omega}t} \oplus {\rm e}^{-i {\bf \Omega}t}) 
(\hat z_0' - z'_\star), 
\end{eqnarray}
where 
${\bf \Omega} := 
{\rm Diag}(\mu_1,...,\mu_n)$, 
the numbers $\mu_k$ are the symplectic eigenvalues of ${\bf H}$, 
and 
\begin{equation}
\hat z_\star' = 
- \tilde{\mathbf S}_{\bf H}^{-\dag}\tilde {\bf H}^{-1}\zeta 
= - {\bf W} {\mathsf S}_{\bf H}^{-\top}{\bf H}^{-1}\xi
=- {\bf W} x'_\star, 
\end{equation}
for $x'_\star$ defined below Eq.~(\ref{Eq:NmHam}). 
At the end, the evolution of the quantum normal modes 
is the quantization of Eq.~(\ref{Eq:CompSol}). 

The solution written in Eq.~(\ref{Eq:CompDiagSol}) only depends on the symplectic spectrum, 
which is invariant under real symplectic transformations. 
In particular, 
${\bf \Lambda}_{{\sf Z}{\bf H}{\sf Z}} = {\bf \Lambda}_{\bf H}$, 
and there is no need to bother with $\mathsf Z$ in Eq.~(\ref{Eq:ComplRep}). 
Note also that the symplectic spectrum, see Eq.~(\ref{Eq:tw3}), 
can be obtained directly from the Euclidean spectrum of ${\sf J}\tilde{\bf H}$,
since $\det({\sf J}\tilde{\bf H} - \lambda {\bf I}_{2n}) = 
\det({\sf J}{\bf H} - \lambda {\bf I}_{2n})$, which follows from  
${\mathbf W} \!\mathsf J {\mathbf W} = \mathsf J$ and $\det {\bf W} = 1$.

\section{Statistical Mechanics}  \label{Sec:SM}                           
The state of a physical system when it attains the equilibrium with a thermal reservoir 
at absolute temperature $T$ is described by the 
{\it canonical density operator}\upcite{landau2,huang,pathria} 
\begin{equation}\label{Eq:thstate}
\hat \rho_\text{T} = \frac{{\rm e}^{-\beta \hat h}}{\mathcal Z_{\beta}}, \,\,\, 
\mathcal Z_{\beta} := {\rm Tr}\, {\rm e}^{-\beta \hat h}, 
\end{equation}
where $\beta:= (k_{\rm B} T)^{-1} \in \mathbb R$ is the ``inverse temperature'', 
$k_{\rm B}$ is the Boltzmann constant and $\hat h$ is the Hamiltonian of the system. 
The partition function $\mathcal Z_{\beta}$ 
provides the normalization of the state in the sense that 
${\rm Tr} \hat \rho_{\rm T} = 1$.

Consider a quadratic Hamiltonian 
$\hat h = H(\hat x)$ for the function $H$ in Eq.~(\ref{Eq:ClQuadLinHam}).  
As learnt in previous sections, 
the condition ${\bf H} > 0$ ensures that the dynamics of a system describes 
a collection of harmonic oscillators in appropriate coordinates.
In statistical physics\upcite{landau2,huang,pathria} 
it is customary to deal with the equilibrium properties 
of these systems in the language of creation-annihilation operators.     
To this end, the transformation 
\begin{equation}\label{Eq:NmTransf}
\hat z =   {\mathbf W} {\mathsf S}_{\bf H}^{-\top} \left(  \hat x + 
           {\bf H}^{-1} \xi \right),  
\end{equation}
which is the composition of the complexification in Eq.~(\ref{Eq:ComplRep}) 
with ${\mathsf L} = {\bf I}_{2n}$ and the affine symplectic coordinate change 
in Eq.~(\ref{Eq:affineTrans}), will be applied to the system Hamiltonian.  
Indeed, 
\begin{eqnarray}\label{Eq:DiagHam}
\tilde H(\hat z) :\!&=&  
H( {\mathsf S}_{\bf H}^{\top} {\bf W}^\ast \hat z - {\bf H}^{-1} \xi )  \nonumber \\  
&=&\tfrac{1}{2} 
\hat z^\dag {\mathbf W}{\bf \Lambda}_{\bf H} {\mathbf W}^\ast \hat z + H'_0 \nonumber \\  
&=& \sum_{k=1}^n \hbar \mu_k (\hat a_k^\dag \hat a_k + \tfrac{1}{2}) 
+ H'_0,   
\end{eqnarray}
where $H'_0 = - \tfrac{1}{2}\xi\cdot {\bf H}^{-1} \xi + H_0$ 
is the same constant as in Eq.~(\ref{Eq:NmHam}). 

The partition function Eq.~(\ref{Eq:thstate}) thus becomes 
\begin{eqnarray} \label{Eq:PartFuncCalc}
\mathcal Z_{\beta} &=& {\rm Tr}\, {\exp}[-\beta H(\hat x)]  = 
{\rm Tr}\, {\exp}[-\beta \tilde H(\hat z)]\nonumber \\ 
&=&{\rm e}^{- \beta H'_0 } \,  
{\rm Tr}\, {\exp}\!\left[- \beta 
\sum_{k=1}^n \hbar \mu_k (\hat a_k^\dag \hat a_k + \tfrac{1}{2})  \right] \nonumber \\ 
&=&{\rm e}^{- \beta H'_0} \prod_{k =1}^n  Z_k, 
%
\end{eqnarray}
where $Z_k$ is the partition function of one oscillator\upcite{landau2,huang,pathria}: 
\begin{equation}
Z_k = {\rm Tr}\, {\exp}\!
 \left[ -\beta\hbar \mu_k 
         (\hat a_k^\dag \hat a_k + \tfrac{1}{2})  \right]
= \tfrac{1}{2}{\rm csch}(\tfrac{1}{2} \beta\hbar \mu_k).
\end{equation}
Consequently, 
\begin{equation}\label{Eq:PartFuncWT}
\mathcal Z_{\beta} 
= \frac{{\rm e}^{- \beta H'_0 }}{2^n} 
\prod_{k =1}^n {\rm csch}\left(\tfrac{1}{2} \beta\hbar \mu_k\right). 
\end{equation}
Finally, the thermal state Eq.~(\ref{Eq:thstate}), 
using Eqs.~(\ref{Eq:DiagHam}) and (\ref{Eq:PartFuncWT}) becomes 
%

\begin{equation}\label{Eq:ThstateWT}
\hat \rho_{\rm T} = 
\hat \rho_{\rm T}^{(1)} 
\otimes ... \otimes \hat \rho_{\rm T}^{(n)}, \,\,\,  
\hat \rho_{\rm T}^{(j)} := 
\frac{ {\rm e}^{ -\beta\hbar \mu_j 
(\hat a_j^\dag \hat a_j + \tfrac{1}{2})}}
{ \frac{1}{2} {\rm csch}\left(\tfrac{1}{2} \beta\hbar \mu_j\right) }.  
\end{equation} 

By virtue of the Williamson theorem, the partition function Eq.~(\ref{Eq:PartFuncWT}) 
is written only in terms of the symplectic spectrum of 
the Hessian of the Hamiltonian, becoming an invariant quantity under 
symplectic transformations 
due to the natural invariance of the symplectic spectrum. 
As is clear in this equation, this theorem also reduces the partition function
of the original system to the one of 
a collection of independent harmonic oscillators.  
The transformation in Eq.~(\ref{Eq:NmTransf}) moves the system to 
the normal-mode coordinates, where the eigenfrequencies are the 
symplectic eigenvalues. 

The internal energy (or simply energy) of a thermodynamical system in equilibrium is 
the mean value of the Hamiltonian: 
$ U: = \langle \hat h \rangle = {\rm Tr}( \hat h \hat \rho_{\rm T})$. 
A system is said to be {\it thermodynamically stable} 
if addition (subtraction) of heat on the system never decreases (increases) 
its temperature. 
Physically speaking, it is a very reasonable and intuitive property, 
since its violation implies that the system will never attain 
an equilibrium state with any other system or with a thermal bath.      
Mathematically, the thermal stability of matter is represented by the positivity 
of the {\it heat capacity}\upcite{landau2,huang,pathria}, which is proportional to the ratio 
of the injected heat and the variation of the temperature.   
For a system in the state Eq.~(\ref{Eq:thstate}), it is given by\upcite{landau2,huang,pathria} 
\begin{equation}
\begin{aligned}
C &= \frac{\partial U}{\partial T} = k_{\rm B} \beta^2 
\frac{\partial^2 }{\partial \beta^2}\! \ln {\mathcal Z_{\beta}} \\
&=  \sum_{k =1}^n \frac{\hbar^2\mu_k^2}{k_{\rm B} T^2} \, 
{\rm csch}^2\!\left(\frac{\hbar \mu_k}{2 k_{\rm B} T} \right),  
\end{aligned}
\end{equation}
where the last equality was obtained using 
the partition function in Eq.~(\ref{Eq:PartFuncWT}).
Consequently, all the Hamiltonians with a 
positive-definite Hessian are thermodynamically stable. 
Thermodynamical instability does occur; 
examples of systems presenting this 
anomalous behavior are discussed in Ref.~\onlinecite{NegHeatCap}.
For quadratic Hamiltonians, 
the simplest example would be a negative definite Hessian, 
where the convergence of the trace in Eq.~(\ref{Eq:PartFuncCalc}) would not happen;  
other examples for the divergence of the partition function 
in the quadratic scenario are analyzed in Ref. \onlinecite{nicacio16}.

The invariance of the partition function under symplectic transformations 
is directly extended for all the thermodynamical functions that are derived from it. 
For instance, 
the internal energy can be written as 
$U := - \frac{\partial}{\partial \beta} \ln {\mathcal Z_\beta}$, 
the Helmholtz free energy of the system is 
$F := - k_{\rm B} T \ln \mathcal Z_\beta$, 
and the entropy $ S =  k_{\rm B}\beta (U-F)$. 
Of course, the above heat capacity is also invariant.
These are highly nontrivial conclusions and were only possible 
due to the Williamson theorem:  
at a first glance, two symplectically congruent Hamiltonians 
may appear very distinct from each other, 
however the thermodynamical behavior of the system will be the same since 
it only depends on the symplectic spectrum. 


When the zero-point energy of the higher frequency oscillator 
is small compared to the thermal energy, 
$\hbar \beta \mu_n = \hbar \mu_n/(k_{\rm B}T) \ll 1$, 
the classical limit is attained by the expansion of 
Eq.~(\ref{Eq:PartFuncWT}) in powers of $(\hbar \beta \mu_k)$ 
up to first order:
\begin{equation}
\mathcal{Z}_{\beta} \longrightarrow \mathcal{Z}_{\beta}^{\rm c} = 
\frac{(k_{\rm B}T)^n {\rm e}^{-\beta H'_0} }
     {\hbar^n \prod_{k=1}^n\mu_k} = 
%
%
\frac{ {\rm e}^{ - \beta H_0 + \tfrac{\beta}{2}\xi\cdot {\bf H}^{-1} \xi  }}
     { (\hbar\beta)^n  \sqrt{\det {\bf H}} } \, . 
\end{equation}
This limit is the classical partition function 
\begin{equation}
\mathcal Z_{\beta}^{\rm c} := 
\frac{1}{(2\pi\hbar)^{n}}\int_{\mathbb R^{2n}} \!\!
{\rm d}^{2n}x \,\, {\rm e}^{-\beta H(x)} 
%
\end{equation}
of the classical Hamiltonian in Eq.~(\ref{Eq:ClQuadLinHam}) with ${\bf H} > 0$. 
The above Gaussian integral is promptly performed 
after the canonical transformation in Eq.~(\ref{Eq:affineTrans}).    
As in the quantum case, all thermodynamical functions will only
depend on the symplectic spectrum and will be also symplectically invariant. 
%

\section{Uncertainty principle} \label{Sec:UP}       
In quantum mechanics, 
noncompatible observables --- 
the ones represented by noncommu\-ting operators ---  
can not be determined with unlimited precision.  
This is a consequence of uncertainty relations. 
In this section, after some words about uncertainty relations, 
the application of the Williamson theorem in this new scenario will be performed 
to reveal invariant structures common to all physical states. 

If a physical system is described by the state $|\psi\rangle \in \mathcal H$, 
the mean-value of an operator $\hat A$ in such state 
is defined by 
$
\langle \hat A \rangle := \langle \psi | \hat A | \psi \rangle.
$
Defining also a {\it displaced observable} as 
$
\Delta \hat A := \hat A - \langle \hat A \rangle, 
$
the variance of measurements of $\hat A$ is expressed as 
\begin{equation}\label{Eq:variance}
\langle \Delta \! \hat A ^2\rangle 
= \langle (\hat A - \langle \hat A \rangle)^2 \rangle =                
\langle \hat A^2 \rangle - \langle \hat A \rangle^2 \ge 0. 
\end{equation}
If another operator, say $\hat B$, is considered,
measurements in the same state 
are constrained\upcite{sakurai,ballentine,cohen} by
\begin{equation}\label{Eq:RobUR}
\langle \Delta \! \hat A^2 \rangle\langle\Delta \! \hat B^2  \rangle \ge 
\tfrac{1}{4} | \langle [\Delta\!\hat A, \Delta\!\hat B ] \rangle |^2
+
\tfrac{1}{4} | \langle \{\Delta\!\hat A, \Delta\!\hat B  \} \rangle|^2,
\end{equation}
where $\{ \hat A, \hat B \} := \hat A\hat B + \hat B\! \hat A$. 
Relation (\ref{Eq:RobUR}), first derived by E. Schrödinger\upcite{Schro},  
is a sufficient condition to the Robertson inequality\upcite{Rob}
\begin{equation}\label{Eq:HeisUR}
\langle \Delta \! \hat A^2 \rangle\langle\Delta \! \hat B^2  \rangle \ge 
\tfrac{1}{4} | \langle [\Delta\!\hat A, \Delta\!\hat B ] \rangle |^2,  
\end{equation}
since $| \langle \{\Delta\!\hat A, \Delta\!\hat B  \} \rangle| \ge 0$. 
This inequality and the one in Eq.~(\ref{Eq:RobUR}) are valid for 
any two operators. Specially when these operators are position and momentum,
Eq.~(\ref{Eq:HeisUR}) receives the name of Heisenberg\upcite{Rob,Schro}. 
For a one-degree-of-freedom system, labeled by $j$, 
the commutation relation is $[\hat q_j,\hat p_j] = i\hbar$ and, 
from Eq.~(\ref{Eq:HeisUR}), the Heisenberg uncertainty principle is written as  
\begin{equation}\label{Eq:HeisUR1D}
\begin{aligned}
\Xi_j \! &:=  
\langle  \Delta \hat q_j^2  \rangle \langle  \Delta \hat p_j^2  \rangle - 
\displaystyle{\frac{\hbar^2}{4}}\ge 0. 
\end{aligned}
\end{equation}
%
%

For $n$ independent systems or a system of 
$n$ noninteracting degrees of freedom, each pair coordinate-momentum will 
obey an inequality in Eq.~(\ref{Eq:RobUR}), 
or its weaker form Eq.~(\ref{Eq:HeisUR}), 
that is $\Xi_j \ge 0$ for $j = 1,...,n$. 
However, if the systems or the degrees of freedom are interacting, 
certainly there will be other correlations (covariances) such as  
$\Delta \hat q_j \Delta \hat q_k$, 
$\Delta \hat q_j \Delta \hat p_k$, or 
$\Delta \hat p_j \Delta \hat p_k$, 
which are not taken into account by Eq.~(\ref{Eq:HeisUR1D}). 
For these remaining pairs of observables, 
other uncertainty relations can be derived from Eq.~(\ref{Eq:RobUR}), 
summing up $n(2n +1)$ dependent inequalities\upcite{Note:NumIneq}.
Thinking in a practical situation, 
if one possesses a set of data corresponding to 
mean-values, variances, and covariances of a system, 
the number of inequalities grows quadratically with $n$. 
The Williamson theorem shows again a way to treat the cases for 
a generic number of degrees of freedom. 

%

To this end, an uncertainty relation taking 
into account all the covariances of the system 
and generalized for mixed states will be constructed. 
Afterwards, a symplectic diagonalization will be performed through the Williamson theorem 
to determine the invariant characteristics of this uncertainty relation.  
The results within the next subsections were originally 
reported in Refs.\onlinecite{simon1994} and \onlinecite{narcowich}, 
while the derivation of the generalized uncertainty relation, 
despite being inspired by the same works, follows a proper pedagogical way.  

\subsection{Robertson-Schrödinger Uncertainty Relation} 
In general, the state of a quantum system is mixed and described 
by a density operator\upcite{sakurai,ballentine,cohen,landau2,huang,pathria} 
$\hat \rho \in \mathcal H\otimes \mathcal H^\dag $, 
where $\mathcal H$ is the Hilbert space of the system. 
The mean value of observables are calculated through 
$\langle \hat A \rangle := {\rm Tr}(\hat \rho \hat A )$
%
and the pure state case is recovered when 
$\hat \rho = |\psi \rangle \! \langle \psi |$. 

Writing as before $\Delta \hat x_j = \hat x_j-\langle \hat x_j\rangle$ 
and using the commutator and the anti-commutator definitions, 
the identity 
\begin{equation}
\tfrac{1}{2}\{\Delta \hat x_j, \Delta\hat x_k\} + 
\tfrac{1}{2}[\Delta \hat x_j, \Delta\hat x_k] = 
\Delta\hat x_j \Delta\hat x_k
\end{equation}
is trivially constructed. Using the commutation relation Eq.~(\ref{Eq:CCR}) 
and taking its mean value, this identity is rewritten as
\begin{equation}\label{Eq:DerUR1}
{\bf V} +\frac{i\hbar}{2} {\mathsf J} = 
\langle \Delta\hat x \Delta\hat x^\top \rangle,  
\end{equation}
where $\bf V$ is the {\it covariance matrix} of the system, 
defined through the matrix elements 
\begin{equation}\label{Eq:DerUR2}
\!{\bf V}_{\!jk} := 
\tfrac{1}{2}\langle \{\Delta \hat x_j, \Delta\hat x_k\} \rangle, \,\,
\end{equation}
and $\langle \Delta\hat x \Delta\hat x^\top \rangle \! 
\in \! {\rm M}(2n, \mathbb R)$ is the matrix with elements\upcite{Note:Notation} 
$\langle \Delta\hat x \Delta\hat x^\top \rangle_{jk} := 
\langle \Delta\hat x_j \Delta\hat x_k \rangle$.  
 
The next step towards the derivation of the new uncertainty relation 
is to prove that 
\begin{equation}\label{Eq:Posit}
\langle \Delta\hat x \Delta\hat x^\top \rangle \ge 0,
\end{equation}
which is performed in the Supplementary Material\upcite{SupMatPosit}.
Finally, the matrix version of the uncertainty relation is composed 
joining Eqs.~(\ref{Eq:DerUR1}), (\ref{Eq:DerUR2}) and (\ref{Eq:Posit}):
\begin{equation}\label{Eq:UR}
{\bf \Delta}:= {\bf V} +\frac{i\hbar}{2} {\mathsf J} \ge 0,    
\end{equation}
which means that $\bf \Delta$ is a Hermitian positive-semidefinite matrix. 
The covariance matrix, 
due solely by the commutation relation in Eq.~(\ref{Eq:CCR}), 
is constrained to such uncertainty relation. 
%
%

For a diagonal covariance matrix, 
\begin{equation}
{\bf V} = 
{\rm Diag}(\langle \Delta \hat q_1^2 \rangle,...,\langle \Delta \hat q_n^2 \rangle,
           \langle \Delta \hat p_1^2 \rangle,...,\langle \Delta \hat p_n^2 \rangle ), 
\end{equation}  
the uncertainty relation in Eq.~(\ref{Eq:UR}) can be easily stated in terms 
of the Euclidean eigenvalues of the matrix $\bf \Delta$. 
These eigenvalues are given by  
\begin{equation}
\begin{aligned}
\delta_j^{\pm} = & - \tfrac{1}{2}
(\langle \Delta \hat q_j^2  \rangle + \langle \Delta \hat p_j^2  \rangle)  \\
&\pm 
\tfrac{1}{2}\sqrt{{(\langle \Delta \hat q_j^2  \rangle + 
 \langle \Delta \hat p_j^2  \rangle  )^2} - \Xi_j }\, ,  
\end{aligned}
\end{equation}
where $\Xi_j$ is the quantity in Eq.~(\ref{Eq:HeisUR}) and $j=1,...,n$. 
The matrix $\bf \Delta$ will be positive semidefinite if and only if 
$\delta_j^+ \ge 0, \, \delta_j^- \ge 0, \forall j$, 
which reduces exactly to $n$ conditions $\Xi_j \ge 0$ in Eq.~(\ref{Eq:HeisUR1D}).
This shows the equivalence of the uncertainty relation Eq.~(\ref{Eq:UR}) with 
$n$ uncertainty relations for noninteracting 
degrees of freedom of the form Eq.~(\ref{Eq:HeisUR1D}).
Remember, however, that Eq.~(\ref{Eq:UR}) is defined for any mixed state, 
while the uncertainty relation Eq.~(\ref{Eq:HeisUR1D}) is written only for pure states.  

\subsection{Williamson Theorem and Symplectic Invariance} \label{Sec:WtSI} 
The covariance matrix in Eq.~(\ref{Eq:DerUR2}) can be rewritten as 
\begin{equation}
{\bf V} = \tfrac{1}{2} \langle 
\Delta\hat x \Delta\hat x^\top +  
(\Delta\hat x \Delta\hat x^\top)^\top \rangle,
\end{equation}
which 
is a sum of two positive semidefinite matrices from Eq.~(\ref{Eq:Posit}), 
thus ${\bf V} \ge 0$. 
Consequently, ${\bf V} > 0$ if and only if $\det {\bf V} \ne 0$. 
In this case, by the Williamson theorem, it is possible to write 
${\sf S}_{\bf V} {\bf V} {\sf S}_{\bf V}^\top = {\bf \Lambda}_{\bf V}$ 
and attain, from the uncertainty relation Eq.~(\ref{Eq:UR}), that
\begin{equation}\label{Eq:URWT}
{\bf \Delta}' := {\sf S}_{\bf V} {\bf \Delta} {\sf S}_{\bf V}^\top = 
{\bf \Lambda}_{\bf V} +\frac{i\hbar}{2} {\mathsf J} \ge 0, 
\end{equation}
since ${\sf S}_{\bf V} {\sf J} {\sf S}_{\bf V}^\top = {\sf J}$. 
Due to the fact that ${\bf \Lambda}_{\bf V}$ is diagonal, 
using the formula for the determinant of block matrices in Sec.\ref{Sec:wt}, 
it is easy to find the $2n$ {\it Euclidean} eigenvalues of the matrix $\bf \Delta'$: 
\begin{equation}
\delta_j'^{\pm} = \mu_j \pm \tfrac{1}{2}\hbar \,\,\,\,\,\,  (j=1,...,n), 
\end{equation}
where $\mu_j$ are the {\it symplectic} eigenvalues of $\bf V$, see Eq.~(\ref{Eq:tw1}). 
%
%
The positive-semidefiniteness of $\bf \Delta'$ in Eq.~(\ref{Eq:URWT}) 
is thus guaranteed if and only if $\delta_j'^{+} \ge 0$ and 
$\delta_j'^{-} \ge 0$, which is equivalent to saying that 
$\mu_j \ge {\hbar}/{2}, \forall j$.
Note that these last conditions subsume the fact ${\bf V} > 0$; 
that is, the positive-definiteness of ${\bf V}$ is automatically 
satisfied for a state such that ${\bf \Delta} \ge 0$.   

The uncertainty relation in Eq.~(\ref{Eq:UR}) can now be rephrased: 
{\it a quantum system has all symplectic eigenvalues 
(of the covariance matrix) greater or equal than $\hbar/2$}.  

The invariance of the commutation relation in Eq.~(\ref{Eq:InvCCR}) 
shows that there is not a preferable set of operators $\hat x$ 
to describe the system. 
Consequently, the uncertainty relation as expressed in terms 
of symplectic eigenvalues is a structural property of the system, 
since the symplectic spectrum is also invariant 
under symplectic transformations.  
Thinking in terms of a symplectic change of coordinates, 
the transformation $\hat x' = {\sf S} \hat x$ for 
$\mathsf S \in {\rm Sp}({2n,\mathbb R})$ turns the covariance matrix,
defined in Eq.~(\ref{Eq:DerUR2}), into
\begin{equation}\label{Eq:CovV}
{\bf V}'_{\!jk} = 
\tfrac{1}{2} \sum_{l,m=1}^{2n} {\sf S}_{jl} {\sf S}_{km}
\langle \{\Delta \hat x'_l, \Delta\hat x'_m\} \rangle
= \left({\sf S} {\bf V}{\sf S}^\top\right)_{_{\!jk}}. 
\end{equation}
Note that ${\bf V}$ and ${\bf V'} = \mathsf S {\bf V} \mathsf S^\top$
share the same symplectic spectrum. 
Defining also 
${\bf \Delta}' := \mathsf S {\bf \Delta} \mathsf S^\top$ for $\bf \Delta$ 
in Eq.~(\ref{Eq:UR}), thus, ${\bf \Delta}' \ge 0$ 
if and only if ${\bf \Delta} \ge 0$, 
which shows that the true important quantity is not 
the covariance matrix itself, 
but its symplectic spectrum.   

If in Eq.~(\ref{Eq:CovV}) $\mathsf S = {\sf S}_{\bf V}$, 
where ${\sf S}_{\bf V} {\bf V} {\sf S}_{\bf V}^\top = {\bf \Lambda}_{\bf V}$, 
the transformation moves the set of system operators to a new set 
where the covariance matrix is ${\bf V}' = {\bf \Lambda}_{\bf V}$.  
In this case, 
both the variances in position and in momentum for 
the same degree of freedom are equal to a symplectic eigenvalue of $\bf V$,  
{\it i.e.}, 
$\langle \Delta \hat q_j'^2 \rangle = 
\langle \Delta \hat p_j'^2 \rangle = \mu_j$.  


At the end, a classical covariance matrix is defined as 
${\bf V}_{\!\rm c} := \tfrac{1}{2} \langle 
\Delta x \Delta x^\top +  
(\Delta x \Delta x^\top)^\top \rangle  =\langle \Delta x \Delta x^\top \rangle 
\in {\rm M}(2n,\mathbb R)$, 
where the mean-values are taken with respect to 
a classical probability density function on phase space\upcite{landau2,huang,pathria}.
Since ${\rm M}(2n,\mathbb R) \ni \Delta x \Delta x^\top \ge 0$, 
thus ${\bf V}_{\!\rm c}\ge 0$. 
Contrary to the quantum case, 
the commutator between classical variables is always null, 
thus ${\bf V}_{\!\rm c}$ is not subjected to any uncertainty relation. 
Consequently, it is possible that $\det {\bf V}_{\!\rm c} = 0$, 
which represents an absolute precision of the measurement of an observable
(a linear combination of positions and momentum), {\it i.e.}, 
the variance of such observable is null. 
If $\det {\bf V}_{\!\rm c} > 0$, 
the Williamson theorem can be applied and the symplectic 
eigenvalues can attain any positive value.  

An example of the uncertainty relation for thermal states is found 
in the Supplementary Material\upcite{SupMatEx}. 

\section{Final Remarks} \label{Sec:FR}   
The widely known Williamson theorem is actually a small piece 
(case $\gamma$ in p.162) of Williamson's original work\upcite{williamson}.     
According to Arnol'd\upcite{arnold},  
D.M. Galin has collected and reinterpreted 
the Williansom results in a classical mechanics point of view, 
which are thus summarized in Appendix 6 of book \onlinecite{arnold}, and deals 
with all the possible normal forms of generic quadratic Hamiltonians.

%
A normal form is understood as the simplest form to which a Hamiltonian is brought by 
symplectic congruences. 
Here, the Ha\-mil\-tonian Eq.~(\ref{Eq:NmHam}) 
is the normal form of Eq.~(\ref{Eq:ClQuadLinHam}).   
In principle, the examples considered in this paper can be extended for more 
generic cases using the list of Galin. 
However, what makes the Williamson theorem useful, practical, and celebrated 
is the particular normal form attained through Eq.~(\ref{Eq:tw1}), 
which only works for positive-definite matrices. 
Although all the other normal forms are no longer diagonal, 
the structure of this paper and the basic concepts using symplectic theory 
serve as a starting point to the treatment of generic cases. 
For instance, 
statistical properties of systems governed by a generic quadratic Hamiltonian 
are described in Ref. \onlinecite{nicacio16} and constitute the generalization 
of the results in Sec.\ref{Sec:SM}. 
Surprisingly enough, not all of these are thermodynamically stable systems; 
however, the thermodynamical properties are symplectically invariant, 
like the stable case analyzed here.  

To the interested reader, a detailed and introductory review 
on the symplectic formalism and its relation with 
quantum mechanics is Ref.~\onlinecite{littlejohn1986}, 
while advanced mathematical background, 
rigorous results, and the state of the art 
are found in Ref.~\onlinecite{gossonbook2006}. 
An enjoyable discussion of nontrivial consequences 
of symplectic geometry in classical and quantum mechanics 
is Ref. \onlinecite{gosson2}. 

The applicability of the Williamson theorem is spread over physics and 
goes far beyond the presented subjects.  
The transformation in Eq.~(\ref{Eq:BogTrans}) 
is a multimode Bogoliubov transformation\upcite{bogoliubov}, 
an ubiquitous method in solid state physics, field theory and quantum optics.  
As a current research area in quantum information, 
entanglement is a genuine quantum property of composite
(in our notation $n\ge 2$)  
and interacting systems.  
A relation almost equal to Eq.~(\ref{Eq:UR}) 
is used to verify its existence\upcite{simon2000}. 
Again the Williamson theorem plays a fundamental 
role and symplectic eigenvalues are used to 
quantify how much a system is entangled\upcite{adesso}.   
The very same procedure presented in Sec.\ref{Sec:QM} is applied to describe 
the propagation of information, heat, classical and quantum correlations 
({\it e.g.} entanglement) through bosonic chains in 
Ref. \onlinecite{nicacio8}. 

The author ultimately hopes that students, 
teachers, and researchers should face the developed subject 
as a new card up their sleeves, 
expanded far beyond the set of examples presented here. 

\acknowledgments        
\noindent The warm hospitality of NuHAG -- 
Universität Wien is acknowledged, 
mainly due to Prof. H.G. Feichtinger, Prof. M. de Gosson, and C. de Gosson. 
I am also grateful for the enthusiasm of Prof. F.L.S. Semi\~ao and 
Prof. C. Farina that encouraged and supported the idea of this work. 
I would like to thank the diligent work of the two anonymous 
referees, which has improved the quality of this article. 
The author is a member of the Brazilian National Institute of Science 
and Technology for Quantum Information [CNPq INCT-IQ (465469/2014-0)] 
and also acknowledges the Brazilian agency CAPES 
[PrInt2019 (88887.468382/2019-00)] by partial financing.


\clearpage
\pagebreak
\setcounter{equation}{0}
\setcounter{figure}{0}
\setcounter{table}{0}
\setcounter{section}{0}
\setcounter{page}{1}
\renewcommand{\thepage}{Supplementary Material-\arabic{page}}
\renewcommand{\thesection}{SM\arabic{section}}
\renewcommand{\theequation}{SM-\arabic{equation}}
\renewcommand{\thefigure}{SM\arabic{figure}}
\renewcommand{\bibnumfmt}[1]{[SM#1]}
\renewcommand{\citenumfont}[1]{SM#1}
\onecolumngrid
\thispagestyle{empty}
\begin{center}
\textbf{\large Supplementary Material on \\
``Williamson theorem in classical, quantum, and statistical physics''}
\end{center}
\begingroup
\addtolength\leftmargini{0.1in}
\begin{quotation} 
\vspace{0.9cm}
\noindent This Supplementary Material contains 
{\bf 1.}~A pedagogical proof for the Williamson Theorem 
(Sec.\ref{Sec:wt} of the main text); 
{\bf 2.}~An extension of the results in Sec.\ref{Sec:HM} of the main text for 
generic quadratic Hamiltonians; 
{\bf 3.}~The Lagrangian treatment of oscillations and comparison with 
the Hamiltonian case; 
{\bf 4.}~The demonstration of Eq.(\ref{Eq:Posit}) in Sec.\ref{Sec:UP} 
of the main text;
{\bf 5.}~Three physical motivated examples for the application of the theorem. 

\vspace{0.1cm}

\noindent Equations here are named as (SM-\#), 
while references for equations in the main text appear as (\#). 
This material contains its own bibliography at the end. 
\vspace{1.cm}
\end{quotation}
\endgroup

\twocolumngrid

\section{Proof of Williamson Theorem}            

Mathematical definitions and properties of some objects in 
the Theorem and in the proof can be found in Sec.\ref{Sec:wt} of the main text.
For convenience, the theorem is reproduced here. 

\vspace{0.2cm}

\noindent {\bf Williamson theorem:} 
{\it 
Let ${\bf M} \in {\rm M}(2n,\mathbb R)$ be symmetric 
and positive definite, 
{\it i.e.}, ${\bf M}^\top = {\bf M} > 0$.  
There exists ${\sf S}_{\bf M} \in {\rm Sp}(2n,\mathbb R)$ 
such that 
\begin{equation} \label{SM:tw1}      
\begin{aligned}
& {\sf S}_{\bf M} {\bf M} {\sf S}_{\bf M}^\top 
= {\bf \Lambda}_{\bf M},                      \\  
& {\bf \Lambda}_{\bf M} := 
{\rm Diag}(\mu_1,...,\mu_n,\mu_1,...,\mu_n)  
\end{aligned}
\end{equation}
with
$ 0 < \mu_j \le \mu_k  \,\,\, \text{for} \,\,\, j \le k$. %
Each $\mu_j$ is such that    
\begin{equation} \label{SM:tw2}
\det({\sf J} {\bf M} \pm i \mu_j {\bf I}_{2n}) = 0 
\,\,\,\,\,\, (j = 1,...,n), 
\end{equation}
and the matrix ${\sf S}_{\bf M}$ admits the decomposition
\begin{equation} \label{SM:tw3}
{\sf S}_{\bf M} = 
\sqrt{\!{\bf \Lambda}_{\bf M}} \, {\bf O} \, \sqrt{\mathbf{M}^{-1}}, 
\end{equation}
where ${\bf O} \in {\rm M}(2n,\mathbb R)$ satisfies 
\begin{equation} \label{SM:tw4}
{\bf O} \, \sqrt{\bf M} \, \mathsf J \, 
\sqrt{\bf M} \, {\bf O}^\top = 
{\bf \Lambda}_{\bf M} \mathsf J , \,\,\,  
\end{equation}
and ${\bf O}^\top = {\bf O}^{-1}$, {\it i.e.}, 
is an orthogonal matrix. }                          \hfill\(\Box\) 

\vspace{0.2cm}

As a useful notation for the proof, 
the set containing all the Euclidean eigenvalues of a matrix ${\bf A}$, 
its spectrum, is denoted by ${\rm Spec}_{\mathbb K}({\bf A})$. 
If all the Euclidean eigenvalues belong to the real set, $\mathbb K = \mathbb R$, 
otherwise $\mathbb K = \mathbb C$. 

\vspace{0.2cm}

\noindent {\bf Proof:} 
Consider a symmetric positive definite matrix ${\bf M} \in {\rm M}(2n,\mathbb R)$. 
The matrix defined by 
$\tilde {\bf M} := \sqrt{\bf M} \mathsf J \sqrt{\bf M} \in {\rm M}(2n,\mathbb R)$, 
with 
\begin{equation} \label{SM:JMat}
{\sf J} := 
          \begin{pmatrix}
          {\bf 0}_n   & {\bf I}_n \\
          - {\bf I}_n & {\bf 0}_n 
          \end{pmatrix}               \in {\rm M}(2n,\mathbb R), 
\end{equation}
see Eq.(\ref{Eq:SympCond}), is anti-symmetric 
($\tilde {\bf M}^\top = - \tilde {\bf M}$), 
since $\sqrt{\bf M} = \sqrt{\bf M}^\top$ and $\mathsf J^\top = -\mathsf J$. 
It also has the same eigenvalues of $\mathsf J {\bf M}$, 
since their characteristic polynomials are equal: 
\begin{eqnarray}    \label{SM:spec0}
P(\lambda) :&=& \det(\tilde {\bf M} - \lambda {\mathbf I}_{2n}) \nonumber \\
&=& \sqrt{\det \bf M} \det( {\mathsf J} \sqrt{\bf M} - 
                            \lambda \sqrt{{\bf M}^{-1}} ) \nonumber \\ 
&=& \det({\mathsf J} {\bf M} - \lambda {\mathbf I}_{2n}). 
\end{eqnarray}
Thus, any property of the spectrum of the matrix $\tilde {\bf M}$ 
is shared by the spectrum of ${\mathsf J} {\bf M}$.  

Since $\det \tilde {\bf M} = \det(\mathsf J {\bf M}) = \det {\bf M} \in \mathbb R$, 
complex eigenvalues of $\tilde {\bf M}$ come always in conjugate pairs, 
which is compactly expressed as 
\begin{equation}\label{SM:spec1}
\lambda \in {\rm Spec}_{\mathbb C}(\tilde {\bf M}) 
\Longleftrightarrow
\lambda^\ast \in {\rm Spec}_{\mathbb C}(\tilde {\bf M}). 
\end{equation}

Using again the characteristic polynomial, 
but taking into account the anti-symmetricity of $\tilde {\bf M}$, one has 
\begin{eqnarray}
P(\lambda) &=& \det(\tilde {\bf M} - \lambda {\mathbf I}_{2n}) 
            =  \det(\tilde {\bf M} - \lambda {\mathbf I}_{2n})^\top \nonumber \\
           &=& \det(\tilde {\bf M}^\top - \lambda {\mathbf I}_{2n}) \nonumber \\ 
&=& (-1)^{2n} \det(\tilde {\bf M} + \lambda {\mathbf I}_{2n}), 
\end{eqnarray}
{\it i.e.}, $P(\lambda) = P(-\lambda)$, 
or the eigenvalues come also in symmetric pairs:
\begin{equation}\label{SM:spec2}
\lambda \in {\rm Spec}_{\mathbb C}(\tilde {\bf M})  
\Longleftrightarrow             
-\lambda \in {\rm Spec}_{\mathbb C}(\tilde {\bf M}).  
\end{equation}

If $\lambda \in {\rm Spec}_{\mathbb C}(\tilde {\bf M})$, 
then $\lambda^2$ is an eigenvalue of the matrix $\tilde {\bf M}^2$. 
However, 
$\tilde {\bf M}^2 = \sqrt{\bf M} \mathsf J {\bf M} \mathsf J \sqrt{\bf M}$ 
is a real symmetric matrix, 
thus possessing only real eigenvalues:   
\begin{equation}\label{SM:spec3}
\lambda \in {\rm Spec}_{\mathbb C}(\tilde {\bf M})  
\Longrightarrow             
\lambda^2 \in {\rm Spec}_{\mathbb R}(\tilde {\bf M}^2) 
\subseteq {\mathbb R}.  
\end{equation}

Taking together conditions (\ref{SM:spec1}) and (\ref{SM:spec3}),
an eigenvalue of $\tilde {\bf M}$ must be a pure imaginary number: 
\begin{equation}
\lambda \in {\rm Spec}_{\mathbb C}(\tilde {\bf M})  
\Longrightarrow             
\lambda = i \mu,  \,\,\,  \mu \in {\mathbb R}.  
\end{equation}%
Taking into account condition (\ref{SM:spec2}), 
the spectrum of $\tilde{\bf M}$ is 
\begin{equation}\label{SM:spec5}
{\rm Spec}_{\mathbb C}(\tilde {\bf M}) = 
\{ i \mu_1, - i \mu_1,...,i \mu_n, - i \mu_n\},    
\end{equation}
where $\mu_k \in {\mathbb R} \, \forall k$. 
The assertion in (\ref{SM:tw2}) of the theorem is thus proved, 
since Eq.(\ref{SM:spec0}) shows that 
${\rm Spec}_{\mathbb C}({\sf J \bf M}) = {\rm Spec}_{\mathbb C}(\tilde {\bf M})$.  

Returning to the matrix $\tilde {\bf M}^2$, 
its symmetricity also ensures that there exist an 
orthogonal matrix 
${\bf O} \in {\rm M}(2n,\mathbb R)$, ${\bf O}^\top = {\bf O}^{-1}$,
such that 
\begin{equation}\label{SM:wil1}
{\bf O} \tilde {\bf M}^2 {\bf O}^\top = {\bf D}, 
\end{equation}
where ${\bf D}$ is the diagonal matrix containing the real eigenvalues of 
$\tilde {\bf M}^2$; from condition (\ref{SM:spec3}), 
these eigenvalues are the square of the ones in (\ref{SM:spec5}) and  
the columns of the matrix $\bf O$ can be organized such that 
\begin{equation}\label{SM:Diag}
{\bf D} = - \, {\rm Diag}(\mu_1^2,...,\mu_{n}^2,\mu_1^2,...,\mu_{n}^2).
\end{equation}
Explicitly writing 
$\tilde {\bf M}^2 = \sqrt{\bf M} \, \mathsf J {\bf M} \mathsf J \, \sqrt{\bf M} $,  
and rearranging terms in Eq.(\ref{SM:wil1}), one can rewrite it as 
\begin{equation}\label{SM:wil2}
{\sf S}_{\bf M}^{-\top} {\mathsf J} 
{\bf M} 
{\mathsf J} \, {\sf S}_{\bf M}^{-1} 
= - {\bf \Lambda}_{\bf M}, 
\end{equation}
for ${\sf S}_{\bf M}$ in (\ref{SM:tw3}) and ${\bf \Lambda}_{\bf M}$ in (\ref{SM:tw1}).
Now, assuming that ${\sf S}_{\bf M}$ is a symplectic matrix, see Eq.(\ref{Eq:SympCond}), 
${\sf S}_{\bf M}^{-\top} {\mathsf J} = {\mathsf J} {\sf S}_{\bf M} $, 
Eq.(\ref{SM:wil2}) becomes 
\begin{equation}
{\sf S}_{\bf M}
{\bf M} {\sf S}_{\bf M}^{\top} 
= - {\mathsf J}^\top {\bf \Lambda}_{\bf M} {\mathsf J}^\top = {\bf \Lambda}_{\bf M}, 
\end{equation}
which proves Eq.(\ref{SM:tw1}). 
However, 
it is still necessary to prove that ${\sf S}_{\bf M}$ 
is a symplectic matrix if and only if the matrix $\bf O$ satisfies Eq.(\ref{SM:tw4}), 
which goes as follows. 

From the symplectic condition for ${\sf S}_{\bf M}$ written as Eq.(\ref{SM:tw3}), and noting that 
$[\sqrt{\!{\bf \Lambda}_{\bf M}}, {\mathsf J }] = 
[{{\bf \Lambda}_{\bf M}}, {\mathsf J }] = 0$, 
one obtains 
\begin{equation*}
\begin{aligned}
{\sf S}_{\bf M}^\top {\mathsf J } {\sf S}_{\bf M} = \mathsf J 
&\Longleftrightarrow 
\sqrt{\mathbf{M}^{-1}} \, {\bf O}^\top \sqrt{\!{\bf \Lambda}_{\bf M}}\, 
{\mathsf J }
\sqrt{\!{\bf \Lambda}_{\bf M}} \, {\bf O} \, \sqrt{\mathbf{M}^{-1}} = \mathsf J \\
& \Longleftrightarrow  
{\bf O}^\top {{\bf \Lambda}_{\bf M}} {\mathsf J }\, {\bf O} = 
 \sqrt{\mathbf{M}} \, \mathsf J \sqrt{\mathbf{M}} \,\\
& \Longleftrightarrow  
{{\bf \Lambda}_{\bf M}} {\mathsf J } = 
{\bf O}  \sqrt{\mathbf{M}} \, \mathsf J  \sqrt{\mathbf{M}} \, {\bf O}^\top, 
\end{aligned}
\end{equation*}
which is precisely Eq.(\ref{SM:tw4}). 
Note that the matrix $\bf O$ satisfying Eq.(\ref{SM:tw4}) 
also satisfies Eq.(\ref{SM:wil1}), since the last can be rewritten as 
\begin{equation}
({\bf O} \tilde {\bf M} {\bf O}^\top)({\bf O} \tilde {\bf M} {\bf O}^\top) 
= (\mathsf J {\bf \Lambda}_{\bf M})^2 =  - {\bf \Lambda}_{\bf M}^2 = {\bf D},     
\end{equation} 
consequently, this is the matrix composed by the orthonormal eigenvectors, 
respectively associated to the eigenvalues in (\ref{SM:Diag}),
of the symmetric matrix $\tilde {\bf M}^2$. 
Note also that the matrix 
${\bf \Gamma} := {\bf O} \tilde {\bf M} {\bf O}^\top  - {\mathsf J} {\bf \Lambda}_{\bf M} 
= {\bf O} \sqrt{\bf M} \, \mathsf J \sqrt{\bf M} \,{\bf O}^\top  
- {\mathsf J} {\bf \Lambda}_{\bf M} \in {\rm M}(2n,\mathbb R)$ 
is antisymmetric, ${\bf \Gamma}^\top = -{\bf \Gamma}$, 
thus ${\bf \Gamma} = {\bf 0}_{2n}$ is a system of $n(2n-1)$ independent equations, 
which can be solved for the matrix elements of $\bf O$. 
Since $\bf O$ is orthogonal, it has $n(2n-1)$ independent matrix elements, 
and thus the system of equations can be solved for these unknowns.

%
%

It only remains to prove that $\mu_k > 0, \forall k$. 
Since ${\bf M}$ is positive-definite and is related to ${\bf \Lambda}_{\bf M}$ 
through a congruence, ${\sf S}_{\bf M} {\bf M} {\sf S}_{\bf M}^\top$ thus, 
${\bf \Lambda}_{\bf M}$ is also positive definite, 
and the theorem is proved. \hfill\(\Box\)

\section{Generic Quadratic Hamiltonians}
The state of a mechanical system 
with $n$ degrees of freedom is described by 
a point in the $2n$-dimensional phase-space 
${\mathbb R}^{n}\times {\mathbb R}^{n}$ and 
the Hamiltonian of the system is, 
in principle, a generic smooth function
\begin{equation}
H: \mathbb R^{n} \times \mathbb R^{n} 
                 \times \mathbb R 
                 \longrightarrow \mathbb R  
: (q,p,t) \longmapsto h.  
\end{equation}
A mere rearrangement of the usual Hamilton equations,  
\begin{equation}
\dot q_j =   \partial h/\partial p_j, \,\,\, 
\dot p_j = - \partial h/\partial q_j, 
\end{equation}
attains\textsuperscript{\citenum{SMClassMec}} the compact form
\begin{equation}
\dot x_k = \sum_{l = 1}^n 
{\mathsf J}_{kl} 
\frac{\partial h}{\partial x_l},  
\end{equation}
for a column vector $x \in \mathbb R^{2n}$ and 
the matrix $\mathsf J$ in (\ref{SM:JMat}).

From the theory of ordinary differential 
equations\textsuperscript{\citenum{SMarnoldODE}},  
$\dot x = {\mathsf J} \mathbf H \, x + {\mathsf J} \xi$ 
[Eq.(\ref{Eq:EqMotQuadHam})]  
is a first order nonhomogeneous linear equation with constant coefficients.  
Its solution is expressed by matrix exponentiation. 
The exponential of a matrix ${\bf A} \in {\rm M}(n,\mathbb R)$ 
is defined by the Taylor series: 
\begin{equation}
\sum_{k=0}^\infty {\bf A}^k/k! =: {\exp}({\bf A}) 
\in {\rm M}(n,\mathbb R). 
\end{equation}

For a generic matrix $\bf H$, the solution of Eq.(\ref{Eq:EqMotQuadHam}) is 
\begin{equation}
x(t) = \mathsf S_t x_0 + 
       \int_0^t \!\! {\rm d}\tau \, {\mathsf S}_\tau {\mathsf J}\xi \,\, , 
\end{equation}  
with 
\begin{equation}
\mathsf S_t := {\exp}[{\mathsf J {\bf H} t}] \in {\rm Sp}(2n,\mathbb R),  
\end{equation}
and can be checked by direct substitution. 

For a nonsingular $\bf H$, 
which is the case when ${\bf H} > 0$,   
the above integral can be explicitly performed, 
\begin{equation}
\int_0^t \!\! {\rm d}\tau \, {\mathsf S}_\tau = 
({\mathsf S}_t - {\mathbf I}_{2n})(\mathsf J \mathbf H)^{-1}, 
\end{equation}
and solution (\ref{Eq:FluxQuadLinHam}) is attained. 

\section{Oscillations in Lagrangian Mechanics}                          
The treatment of oscillations traditionally\textsuperscript{\citenum{SMClassMec}}
departs from a Lagrangian function and consists of an expansion around a critical point 
of the potential energy of the system, leading to an approximated Lagrangian 
of the form
\begin{equation} \label{SM:Lagragian}
L(q,\dot q) = \tfrac{1}{2}\dot q \cdot {\bf T} \dot q - 
\tfrac{1}{2} q \cdot {\bf U} q, 
\end{equation}
where $q = (q_1,...,q_n)^\top$ is the vector of the generalized coordinates, 
$\dot q = d q/dt$ are the generalized velocities, 
and ${\bf T},{\bf U}$ are real symmetric matrices.  
The standard recipe\textsuperscript{\citenum{SMClassMec}} 
follows a long procedure to simultaneously diagonalize 
the matrices ${\bf T}$ and ${\bf U}$, 
attaining a Lagrangian of oscillators 
if both ${\bf T} >0$ and ${\bf U} >0$.  
In order to compare with the Hamiltonian treatment presented so far, 
a straightforward Lagrangian approach will be developed.

If ${\bf T} >0$ and ${\bf U} >0$, 
it is possible to define the symmetric matrix 
$\tilde {\bf U}:=
{\sqrt{\bf T}}^{-1} {{\bf U}} {\sqrt{\bf T}}^{-1}$, 
which is a congruence of the matrix $\bf U$, 
thus also positive-definite, $\tilde {\bf U} > 0$. 
Consider now, the orthogonal matrix $\tilde {\bf O}$ 
that diagonalizes $\tilde {\bf U}$,
\begin{equation}\label{SM:LagDiag}
\tilde {\bf O} \tilde {\bf U} \tilde {\bf O}^\top = 
{\bf \Upsilon} := {\rm Diag}(u_1,...,u_n),
\end{equation}
where $u_1 \le u_2 ... \le u_n$. 
The eigenvalues $u_k$ are the roots of the characteristic polynomial
\begin{equation}\label{SM:LagEig}
\det( \tilde{\bf U}  - \lambda {\mathbf I}_{n}) = 
\det( {\bf U} - \lambda {\bf T})\det{\bf T}^{-1} = 0,            
\end{equation}
and are positive, $u_k > 0, \forall k$, since $\tilde{\bf U} >0$. 

According to the diagonalization of the matrix $\tilde {\bf U}$, 
the (point) transformation 
$q' = \tilde {\bf O} {\sqrt{\bf T}} q$ 
transforms Lagrangian (\ref{SM:Lagragian}) into a new one describing
$n$ independent harmonic oscillators:   
\begin{eqnarray} \label{SM:LagOH}
L'(q',\dot q') = \, 
&\tfrac{1}{2}&{\sqrt{\bf T}}^{-1} \tilde {\bf O}^\top \dot q'\cdot {\bf T} 
{\sqrt{\bf T}}^{-1} \tilde {\bf O}^\top \dot q'  +        \nonumber \\ 
 -\,  &\tfrac{1}{2}&{\sqrt{\bf T}}^{-1} \tilde {\bf O}^\top q'\cdot {\bf U} 
{\sqrt{\bf T}}^{-1} \tilde {\bf O}^\top q'                 \nonumber \\
 = \, &\tfrac{1}{2}& \dot q'\cdot \dot q' - \tfrac{1}{2} q'\cdot {\bf \Upsilon} q',   
\end{eqnarray}
which is the desired result. 

To show the equivalence with the Hamiltonian prescription, 
a Legendre transformation is performed in (\ref{SM:Lagragian}): 
\begin{equation}\label{SM:HamLeg}
H(q,p) := p\cdot \dot q - L(q,\dot q) = \tfrac{1}{2} p\cdot {\bf T}^{-1} p 
                                     + \tfrac{1}{2} q\cdot {\bf U} q,  
\end{equation}
where $ p := {\partial L}/{\partial \dot q} = {\bf T} \dot q$. 
This Hamiltonian can be written as the quadratic form (\ref{Eq:ClQuadLinHam}) with 
\begin{equation}\label{SM:HessLag}
{\bf H} =  \left( \begin{array} {rc} 
                 {\bf U} & {\bf 0}_n  \\
                    {\bf 0}_n  & {\bf T}^{-1}
\end{array} \right) > 0, \,\,\, 
{\mathsf J}{\bf H} =  \left( \begin{array} {rc} 
                             {\bf 0}_n &  {\bf T}^{-1}   \\
                              - {\bf U} & {\bf 0}_n  
\end{array} \right). 
\end{equation}
Noting that  
$\det({\mathsf J} {\bf H} - \lambda {\mathbf I}_{2n}) = 
\det({\bf U} + \lambda^2 {\bf T})\det({\bf T}^{-1})$,
Eq.(\ref{Eq:tw2}) and Eq.(\ref{SM:LagEig}) show that 
$u_k = \mu_k^2$. 
Following (\ref{Eq:NmHam}), 
the normal-mode Hamiltonian is 
\begin{equation}
H'(x') = \tfrac{1}{2} p' \cdot \sqrt{{\bf \Upsilon}}\,  p' + 
         \tfrac{1}{2} q'\cdot \sqrt{{\bf \Upsilon}} \, q', 
\end{equation}
which actually is not the Legendre transformation of Lagrangian (\ref{SM:LagOH}).  
However, the symplectic transformation  
$x'' =  ({\bf \Upsilon}^{{1}/{4}} \oplus {\bf \Upsilon}^{-{1}/{4}}) x'$ 
gives rise to
$H''(x'') = \tfrac{1}{2} p'' \cdot  p'' + 
            \tfrac{1}{2} q''\cdot {\bf \Upsilon} q''$.  
Finally, the Hamiltonian $H''(x')$ is the Legendre transformation of 
Lagrangian (\ref{SM:LagOH}) and, consequently, 
the Williamson theorem (supplied by an extra symplectic transformation) 
provides the results of the standard methods.  

The main advantage of the Hamiltonian description is the symplectic structure 
of phase space, where coordinates and momenta are treated on an equal footing.
While the Lagrangian description departs from a separable form 
$L = T - U$, the Hamiltonian is a generic function of phase-space coordinates, 
not restricted to $T + U$.  
This is clearly manifested by the Hamiltonian (\ref{SM:HamLeg}), 
which is a particular instance of the general quadratic case in 
Eq.(\ref{Eq:ClQuadLinHam}). 

\section{Demonstration of Eq.(\ref{Eq:Posit})} 
The spectral decomposition of the density opera\-tor\textsuperscript{\citenum{SMQuantMech,SMStatPhys}} reads 
\begin{equation}
\hat \rho = \textstyle{\sum_{l}} \, p_l \, | \phi_l \rangle \langle \phi_l |, 
\,\,\, 
\textstyle{\sum_{l}} \, p_l = 1, 
\,\,\, 
0 \le p_l \le 1, \forall l , 
\end{equation}
where $| \phi_l \rangle\in \mathcal H$ are the eigenvectors of $\hat \rho$
associated to the eigenvalues $p_l$.
Employing such a decomposition, one obtains 
\begin{eqnarray}
\langle \Delta\hat x_j \Delta\hat x_k \rangle &=& 
{\rm Tr}(\hat \rho \Delta\hat x_j \Delta\hat x_k) \nonumber \\
&=& \textstyle{\sum_{l}} \, p_l 
\langle \phi_l| \Delta\hat x_j \Delta\hat x_k|  \phi_l \rangle;
\end{eqnarray}
using a completeness relation for a generic complete basis 
$|\psi_m\rangle \in \mathcal H$, last equation becomes 
\begin{eqnarray}
\langle \Delta\hat x_j \Delta\hat x_k \rangle &=& \textstyle{\sum_{l,m}} \, p_l \, 
\langle \phi_l| \Delta\hat x_j|\psi_m\rangle \! 
\langle \psi_m| \Delta\hat x_k|  \phi_l \rangle \nonumber \\  
&=&\textstyle{\sum_{l,m}} \, p_l \, w_{(lm)j} w_{(lm)k}^\ast  \nonumber\\
&=& \textstyle{\sum_{l,m}} \, p_l \, \left[w_{(lm)} w_{(lm)}^\dag\right]_{jk}, 
\end{eqnarray}
where $w_{(lm)j} \!:= 
\!\langle \phi_l | \Delta\hat x_j |\psi_m\rangle \!\in \!\mathbb C$ 
is the component $j$ of the vector  
$w_{(lm)}\!:= \!\langle \phi_l | \Delta\hat x |\psi_m\rangle\! \in\! {\mathbb C}^{2n}$. 
For any $(l,m)$, 
the matrix $w_{(lm)}w_{(lm)}^\dag \!\ge\! 0$, 
consequently it is possible to conclude that 
\begin{equation}\label{Eq:DerUR3}
\langle \Delta\hat x \Delta\hat x^\top \rangle = 
\textstyle{\sum_{l,m}} \, p_l\,  w_{(lm)}w_{(lm)}^\dagger \ge 0, 
\end{equation}
since $p_l \ge 0, \forall l$, as it was to be proved.  

%

\section{Examples}           
Three examples will be presented in this section. 
The objective of the first one is to compare the 
results provided by the Williamson theorem and the diagonalization of the Lagrangian 
function. It is designedly written to be independent of the main body of the text, 
in such a way that the reader would be able to understand the comparison without
technical details. 

The second example considers the process of symplectic diagonalization of 
a nontrivial Hamiltonian, 
which the main objective is to show how to perform in practice its symplectic 
diagonalization. 
Once the symplectic spectrum and the symplectic diagonalizing matrix are obtained, 
the determination of the normal modes of the system, both classical and quantum, 
are immediate, as well as the thermal equilibrium state. 

In the third example, 
the uncertainty relations for thermal states associated 
to quadratic Hamiltonians will be examined, as well as 
the relation between the symplectic spectrum 
of the Hamiltonian and the one for the covariance matrix of the state.   

\subsection{Interacting Trapped Ions}
\begin{figure}[!ht]
\includegraphics[width=6.cm, trim=0 0 0 0]{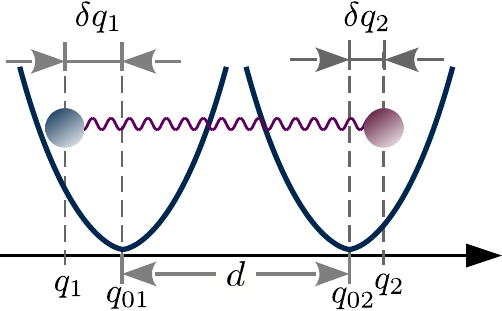}
\caption{(color online) 
Pictorial representation of two interacting trapped ions. 
A parabola represents the confinement 
of one ion to the one-dimensional harmonic motion, 
while the wavelike curve represents the electrical interaction 
between the pair. 
The coordinates of the center of mass of each ion are denoted by 
$q_{j}$, while $q_{0j}$ and $\delta q_j := q_{j} - q_{0j}$ 
are, respectively, 
the equilibrium position (of each trap) 
and the displacement of the equilibrium 
when no electrical interaction (due to the other ion) 
is present. 
The distance between the traps are $d = q_{02} - q_{01}$.
This system is inspired by the work in Ref.\onlinecite{SMnicacio3}.}               \label{Fig01}
\end{figure}

The actual technological scenario is marked by an unprecedented control 
of quantum systems. 
Among them, a single ion is confined inside a trap designed by 
(time-dependent) electromagnetic fields, 
a setup called Paul Trap\textsuperscript{\citenum{SMPaulTrap}} 
in honor of its inventor and Nobel prize awarded.  
This setup combined with laser technics\textsuperscript{\citenum{SMleibfried}} 
is the most developed setup for investigation of quantum effects 
and an imminent candidate for the
construction of a quantum computer\textsuperscript{\citenum{SMzoller}}. 
In a linear trap\textsuperscript{\citenum{SMPaulTrap}}, 
the center of mass of the ion 
is confined to move harmonically in one dimension and, 
since ions are charged 
(usually cations), 
two of them will interact electrically, see Fig.\ref{Fig01}. 

A classical description for the system 
consists of two particles ($j = 1,2$) 
with mass $m$, subjected to a harmonic potential with frequency 
$\varpi$ (the frequency is determined by the trap), 
such that the kinetic energy is 
\begin{equation}\label{Ex:IKen}
T = \frac{m}{2} ({\dot q}_1^2 + {\dot q}_2^2).
\end{equation}
The potential energy of the system, 
taking into account the trapping and 
the electrical interaction, is 
\begin{equation}\label{Ex:IPotU}
U = \frac{m\varpi^2}{2} (\delta q_1^2 + \delta q_2^2)  
+ \frac{C}{|q_1- q_2|},  
\end{equation}
where $C := K_{\rm e} Q_1 Q_2 $ for  
the electrostatic constant $K_{\rm e}$, and  
the ionic charges $Q_1,Q_2$. 

If the distance between the traps 
is much bigger than the displacements of the ions inside the traps, $d \gg \delta q_j$, 
an expansion of the electrostatic potential\textsuperscript{\citenum{SMnicacio3}}
can be performed using $q_1- q_2 = \delta q_1- \delta q_2 - d$, that is
\begin{equation}\label{Ex:Expan}
\begin{aligned}
{|\delta q_1- \delta q_2 - d|^{-1}}  
= \sum_{k =0}^\infty 
\frac{(\delta q_1 - \delta q_2)^k}{d^{k+1}}.  
\end{aligned}
\end{equation}
Keeping only terms up to second order in this expansion, 
the potential energy becomes 
\begin{equation}\label{Ex:QudU}
U \approx \tfrac{1}{2} ({q - q_\star }) 
\cdot {\mathbf U}({q - q_\star}) + U_0, 
\end{equation}
where $q := (q_1, q_2)^\top$ is the column vector of the coordinates and 
the potential matrix is 
\begin{equation}\label{Ex:IPotMat}
\begin{aligned}
{\mathbf U} &:= \left(\begin{array}{cc}
  m \varpi^2 + 2\frac{C}{d^2} &  -2\frac{C}{d^2} \\  
 -2 \frac{C}{d^2}& m\varpi^2 + 2\frac{C}{d^2} 
\end{array} \right).  
\end{aligned}
\end{equation}
The equilibrium coordinate, $q_\star$, 
for the potential energy (\ref{Ex:QudU}) is the solution 
of \textsuperscript{\citenum{SMNoteUniq}} 
\begin{equation}
{\bf U} (q_\star - q_0) = -  \tfrac{C}{d}               
\left(\begin{smallmatrix}
            1 \\
            -1 
      \end{smallmatrix} \right),           
\end{equation}
for $q_0:= (q_{01}, q_{02})^\top$.  
The potential offset is
\begin{equation}
U_0 := C/d -\tfrac{1}{2}(q_\star - q_0)\cdot {\bf U}(q_\star - q_0).
\end{equation}
Note that at $q=q_\star$, $U \approx 0 = T$ and 
the ions are in rest. Note also that $q_\star$ 
is not a critical point of the potential in (\ref{Ex:IPotU}), 
although it might be sufficiently closer for big values of $d$.  

From the kinetic energy in (\ref{Ex:IKen}) and the potential in (\ref{Ex:QudU}), 
the (approximated) Lagrangian of the system is 
\begin{equation}\label{Ex:QuadLag}
L(q,\dot q) = \frac{m}{2} \dot q \cdot \dot q - 
\tfrac{1}{2} ({q - q_\star }) \cdot {\mathbf U}({q - q_\star}) + U_0.  
\end{equation}
One can then perform the point transformation
\begin{equation}
q' = \sqrt{m}\, \tilde{\bf O}(q - q_\star),
\end{equation}
where $\tilde{\bf O}$ is the orthogonal matriz, 
$\tilde{\bf O}^\top = \tilde{\bf O}^{-1}$, that diagonalizes $\bf U$.
Such transformation always exists, 
since the potential matrix in (\ref{Ex:IPotMat}) 
is real and symmetric\textsuperscript{\citenum{SMNotePointT}}.
Indeed, 
\begin{equation}
\tilde{\bf O}{\bf U}\tilde{\bf O}^\top = 
{\rm Diag}\left(m \varpi^2,m \varpi^2 + 4 \frac{C}{d^3}\right). 
\end{equation}
The new Lagrangian becomes
\begin{equation}
\begin{aligned}
L'(q',\dot q') &= 
\tfrac{1}{2} \dot q' \cdot \dot q' - 
\tfrac{1}{2} q' \cdot {\bf \Omega}^2 q'+ U_0 \\
&= 
\tfrac{1}{2}\sum_{k = 1,2} \left(\dot q_k'^2 - \omega_k^2 q_k'^2\right) + U_0,    
\end{aligned}
\end{equation}
where ${\bf \Omega} := {\rm Diag}(\omega_1,\omega_2)$ and 
\begin{equation}
\omega_1 := \varpi, \,\,\, 
\omega_2 := \sqrt{\varpi^2 + \frac{4 C}{m d^3}}. 
\end{equation}
Finally, 
from the Euler-Lagrange equation, one obtains
\begin{equation}\label{Ex:MovEq}
\ddot q_k' + \omega_k^2 \, q_k' = 0\,\,\, (k =1,2).  
\end{equation}

Despite the usual traps deal with cations, 
theoretically it is possible to consider generic charges in (\ref{Ex:IPotU}). 
When the charges of the ions have the same sign ($C>0$), 
then $\omega_k > 0$, and the movement will be oscillatory. 
In this case, $\det {\bf U} > 0$ and the fixed point in (\ref{Ex:QuadLag}) 
is expressed as
\begin{equation}
q_\star = q_0 -\frac{C}{d} {\bf U}^{-1} \left(\begin{smallmatrix}
            1 \\
            -1 
      \end{smallmatrix} \right) 
=
q_0 -\frac{C m\varpi^2}{d \det{\bf U}}  \left(\begin{smallmatrix}
            1 \\
           -1 
      \end{smallmatrix} \right), 
\end{equation}
which means that ion 1 (resp. 2) oscillates 
around a stable equilibrium point translated to the 
left (resp. right) with respect to the center of its 
trap $q_{01}$ (resp. $q_{02}$), 
according to the mutual repulsion of the charges.

In the other case, $C < 0$, 
the ions will attract each other and the movement 
will be stable (oscillatory evolution)
only if $\omega_2^2 > 0$, that is, 
if $\varpi^2 > 4|C|/(m d^3)$. %
The stable fixed $q_\star$ point will 
be displaced in the opposite direction of the previous case, 
due to the attraction. 
On the other hand, if $\varpi^2 < 4|C|/(m d^3)$, 
the equilibrium will be unstable 
since $\omega_2^2 < 0$, thus 
solution $q_2(t)$ in (\ref{Ex:MovEq}) is
such that $\lim_{t \to \infty} q_2(t) = \infty$, 
which means that the trap collapses. 

From the point of view of the Hamiltonian dynamics, 
the Hamiltonian of the original system is 
\begin{equation}\label{Ex:ionHam}
\begin{aligned}
H(q,p) &= \frac{1}{2m} ({\dot p}_1^2 + {\dot p}_2^2) \\ 
&+ \frac{m\varpi^2}{2} (\delta q_1^2 + \delta q_2^2)  
+ \frac{C}{|q_1- q_2|},    
\end{aligned}
\end{equation}
and the same expansion in (\ref{Ex:Expan}) is performed to attain 
the Legendre transform of (\ref{Ex:QuadLag}), which can be written as
\begin{equation}
H(x) = \tfrac{1}{2} (x - x_\star) \cdot {\bf H} (x - x_\star) + U_0,  
\end{equation}
with $ x:=(q_1,q_2,p_1,p_2)^\top$, similarly for $x_\star$,  
and ${\bf H} = m^{-1}\mathbf{I}_{2} \oplus {\bf U}$. 

In the Lagrangian scenario, the movement will be stable if the eigenvalues 
of $\bf U$ are positive, which is the same to say that $\bf U$ is (symmetric)
positive-definite\textsuperscript{\citenum{SMClassMec}}; observe that the positivity character 
of ${\bf U}$ implies that $\bf H$ above is also positive-definite. 
Departing from this fact, the matrix 
$
\mathsf S_{\bf H} := \mathsf L ( {\tilde {\bf O}}^\top \oplus {\tilde {\bf O}} ) 
$ is such that 
\begin{equation}
\begin{aligned}
{\mathsf S}_{\bf H} {\bf H} {\mathsf S}_{\bf H}^\top &=  
\mathsf L ( {\tilde {\bf O}} \oplus {\tilde {\bf O}} ) \, {\bf H}  
( {\tilde {\bf O}}^\top \oplus {\tilde {\bf O}}^\top )\mathsf L^\top \\
&= \mathsf L \left[(m^{-1}{\bf I}_2) \oplus {\rm Diag}(m\omega_1^2,m\omega_2^2)\right] 
\mathsf L\\
&= {\rm Diag}(\omega_1,\omega_2,\omega_1,\omega_2),   
\end{aligned}
\end{equation}
where 
\begin{equation}
\mathsf L := {\rm Diag}\left(\sqrt{m\omega_1},\sqrt{m\omega_2},
                       \frac{1}{\sqrt{m\omega_1}},
                       \frac{1}{\sqrt{m\omega_2}}\right)
\end{equation}                       
and ${\tilde {\bf O}}$ is the same as before.
The matrix $\mathsf S_{\bf H}$ performs a symplectic diagonalization of $\bf H$, 
since $\mathsf S_{\bf H}$ satisfies Eq.(\ref{Eq:SympCond}) and the diagonal matrix
${\bf \Lambda}_{\bf H} :=  {\mathsf S}_{\bf H} {\bf H} {\mathsf S}_{\bf H}^\top$ 
is the symplectic spectrum of $\bf H$. 
Note that ${\mathsf S}_{\bf H}^\top \ne {\mathsf S}_{\bf H}^{-1}$.

From the above diagonalization procedure, 
the affine ca\-no\-ni\-cal transformation
$x' = {\mathsf S}_{\bf H}^{-\top}(x-x_\star)$ brings the Hamiltonian to 
\begin{equation}\label{Ex:DiagHamIon}
\begin{aligned}
H'(x') &= \tfrac{1}{2} x' \cdot {\bf \Lambda}_{\bf H} x' + U_0  \\
&= \tfrac{\omega_1}{2}(p_1'^2 + q_1'^2) + 
   \tfrac{\omega_2}{2}(p_2'^2 + q_2'^2), 
\end{aligned}
\end{equation}
which is the Hamiltonian of two harmonic oscillators.  

Trapped ions are naturally described by quantum theory and 
the quantum description of the problem 
is provided by the (symmetric) quantization of the variables: 
$(q,p) \mapsto (\hat q, \hat p)$. 
In turn, the quantum Hamiltonian has the same functional form 
of (\ref{Ex:ionHam}). After this point, the very same treatment 
is performed and the Hamiltonian of quantum oscillators are obtained by 
the same replacement $(q',p') \mapsto (\hat q', \hat p')$ in (\ref{Ex:DiagHamIon}).

\subsection{Quantized Electromagnetic Field}
Consider a quantum system of three degrees of freedom that evolves governed by the
quadratic Hamiltonian $\hat h = \hat H_0 + \hat H_1 + \hat H_2$ where
\begin{equation}\label{Ex:QuantHam}
\begin{aligned}
\hat H_0 &=  {\hbar\omega}\sum_{j=1}^3 (\hat a_j^\dag \hat a_j + \tfrac{1}{2}), \\
\hat H_1 &= \frac{i\hbar\gamma}{2}\sum_{j=1}^3 (\hat a_j^{\dag 2} - \hat a_j^2), \\ 
\hat H_2 &= -\frac{i\hbar \kappa}{\sqrt{2}} 
           (\hat a_1^{\dag}\hat a_2^{\dag} - \hat a_1\hat a_2 + 
            \hat a_2^{\dag}\hat a_3^{\dag} - \hat a_2\hat a_3).
\end{aligned}
\end{equation}
Despite being a toy model, in principle it can be reproduced in a quantum optics lab.  
The Hamiltonian $\hat H_0$ governs 
the evolution of three noninteracting electromagnetic fields 
(``harmonic oscillators'') with equal frequency $\omega$; 
all the other terms are related to the phenomenon known as
squeezing\textsuperscript{\citenum{SMNoteSqueez}}, 
which can be reproduced experimentally by (nonlinear) interactions of the 
electromagnetic field with crystals\textsuperscript{\citenum{SMQuantOptics}}. 
The Hamiltonian $\hat H_1$ represents the squeezing on each electromagnetic field and 
is known as ``one-mode squeezing'', while the terms in $\hat H_2$ are called ``two-mode squeezing'', since each term acts on pairs, and is responsible 
for the creation of entanglement between these field pairs\textsuperscript{\citenum{SMQuantOptics}}. 

Using transformation (\ref{Eq:zdef}) 
with $m_j = 1, \omega_j = 1,  \forall j$ (in suitable units of the problem), 
the Hamiltonian is rewritten as  
$\hat h = H(\hat x)$ for the function $H$ in (\ref{Eq:ClQuadLinHam}) with 
$\xi = 0$, $H_0 =0$ and 
\begin{equation}\label{Ex:Hess} 
\!\!\!\mathbf H  = \!
\left(\begin{array}{cc}
      \omega {\mathbf I}_{3} & {\mathbf C} \\
      {\mathbf C}          & \omega {\mathbf I}_{3}
      \end{array}
\right),  \,\,\, 
\mathbf C =\!
\left(\!\!\begin{array}{ccc}
       \frac{\gamma}{2} & - \frac{\kappa}{\sqrt{2}} & 0  \\
      - \frac{\kappa}{\sqrt{2}} & \frac{\gamma}{2} & - \frac{\kappa}{\sqrt{2}} \\
      0 & - \frac{\kappa}{\sqrt{2}} & \frac{\gamma}{2} 
\end{array}\!\!
\right). 
\end{equation} 

%

To obtain the normal modes of the system in question, 
it is necessary first to check whether the (symmetric) matrix ${\bf H}$ 
is positive-definite.  
To this end, the Euclidean eigenvalues of ${\bf H}$ are determined by roots of the characteristic polynomial 
$\det({\mathbf H} - \lambda {\mathbf I}_{2n}) = 0$, which are organized on the 
following diagonal matrix 
\begin{equation}\label{Ex:SpecH}
\begin{aligned}
{\bf D} = {\rm Diag}( \, & \omega + \gamma,\, 
                                \omega - \kappa + \gamma,\, 
                                \omega + \kappa + \gamma,    \\
 &\omega - \gamma, \, \omega +\kappa -\gamma, \, \omega - \gamma -\kappa \, ).
\end{aligned}
\end{equation}
Since a symmetric matrix is positive definite if and only if its eigenvalues are 
positive, $\omega > \kappa + \gamma$ is a necessary and sufficient condition for 
the positive definiteness of $\bf H$.  
Considering that this is the case, 
the determination of the normal modes of this system is routed by 
the Williamson theorem. 

The first step now is to determine the symplectic spectrum of $\bf H$ 
following (\ref{SM:tw2}); 
thus, solving for the roots of the characteristic polynomial 
$\det({\mathsf J} {\mathbf H} - \mu {\mathbf I}_{2n}) = 0$, 
one finds 
${\bf \Lambda}_{\bf H} = {\rm Diag}(\mu_1,\mu_2,\mu_3, \mu_1, \mu_2, \mu_3)$, 
where
\begin{equation}\label{Ex:SympEigH}
\begin{aligned}
\mu_1 & = \sqrt{\omega^{2} - \gamma^{2}}, \\ 
\mu_2 &= \sqrt{\omega^{2} - (\kappa-\gamma)^{2}}   ,  \\
\mu_3 & =  \sqrt{\omega^2 - (\kappa+\gamma)^2}, 
\end{aligned}
\end{equation}
which are the eigenfrequencies of the system, 
or the frequency of the normal modes. 

The next step is the determination of the symplectic matrix that 
symplectically diagonalizes ${\bf H}$ as in (\ref{SM:tw3}), 
but for that the square-root 
of ${\bf H}^{-1}$ is needed. 
To calculate this square-root, the Euclidean diagonalization 
of $\bf H$ will be performed.

Consider thus the orthogonal matrix ${\sf O}'$ composed by the 
orthonormal eigenvectors of $\bf H$, which are such that 
\begin{equation}\label{Ex:DiagH}
{\sf O}'{\bf H} {\sf O}'^\top =  {\bf D},
\end{equation}
where $\bf D$ is defined in (\ref{Ex:SpecH}). 
The matrix ${\sf O}'$ can be determined by brute force 
with the help of a symbolic computational program, if necessary, 
however, it is useful to show that it can be decomposed as the product 
of two suitable matrices: 
\begin{equation}\label{Ex:DiagH2}
{\sf O}' = \mathsf R \, ({\bf O}_{{\bf C}} \oplus {\bf O}_{{\bf C}}),
\end{equation}
where
\begin{equation*}
\begin{aligned}                                                                        
\mathsf R := \frac{1}{\sqrt{2}}
\left(
\begin{array}{cc}
 {\mathbf I}_{3} & {\mathbf I}_{3}  \\
-{\mathbf I}_{3} & {\mathbf I}_{3}
\end{array}
\right), \,\,\, 
{\bf O}_{{\bf C}} :=
\left(
\begin{array}{ccc}
\frac{1}{2}  & -\frac{1}{\sqrt{2}} & \frac{1}{2}  \\
\frac{1}{2} & \frac{1}{\sqrt{2}} & \frac{1}{2} \\
-\frac{1}{\sqrt{2}} & 0 & \frac{1}{\sqrt{2}} 
\end{array}
\right). 
\end{aligned}
\end{equation*}
The orthogonal matrix $\bf O_{{\bf C}}$ is the one that performs the 
diagonalization of the symmetric matrix ${\bf C}$ in Eq.(\ref{Ex:Hess}), {\it i.e.},
\begin{equation} 
{\bf O}_{\bf C} \, {\bf C} \, {\bf O}_{\bf C}^\top = 
{\bf D}_{\bf C} := {\rm Diag}(\gamma, \gamma-\kappa,\gamma+\kappa ).
\end{equation}
With this in hand, the diagonalization of $\bf H$ is performed in two steps, 
first by the diagonalization of the blocks $\bf C$, 
and then by applying a rotation $\mathsf R$:  
\begin{equation*}
\begin{aligned}
{\sf O}'{\bf H} {\sf O}'^\top &= 
\mathsf R 
\left(\begin{array}{cc}
\omega {\mathbf I}_{3} & {\bf O}_{{\bf C}} \,{\bf C} \,{\bf O}_{{\bf C}}^\top \\
{\bf O}_{{\bf C}} \,{\bf C} \, {\bf O}_{{\bf C}}^\top  & \omega {\mathbf I}_{3}
                \end{array}
                \right) \mathsf R^\top  \\
%
%
 & =         \left(\begin{array}{cc} 
                \omega {\mathbf I}_{3} + {\bf D}_{\mathbf C} & {\bf 0}_3 \\
                {\bf 0}_3   & \omega {\mathbf I}_{3} - {\bf D}_{\mathbf C}
                \end{array}
                \right)  = {\bf D}. 
\end{aligned}
\end{equation*}
Note that ${\mathsf R}^\top = {\mathsf R}^{-1}$, 
and ${\mathsf R} \in {\rm Sp}(6,\mathbb R)$, 
also note that 
$({\bf O}_{{\bf C}} \oplus {\bf O}_{{\bf C}}) \in {\rm Sp}(6,\mathbb R)$ and, 
consequently, ${\sf O}'$ besides orthogonal is also symplectic. 

The two step procedure in last paragraph only works due to $\bf C^\top = \bf C$. 
As a clue, in practical problems, 
for instance, the ones in Ref.\onlinecite{SMnicacio8},
is common to find a Hamiltonian 
where the blocks can be diagonalized one at a time, 
and thus a final rotation can be used to diagonalize the whole matrix.  
This is the reason to illustrate it here. 
In the absence of this structure, 
or other symmetry like it,  
symbolic computational programs 
solves the problem with efficiency. 

The symplectic matrix, 
which moves the system to nor\-mal-modes coordinates, 
from Eq.(\ref{SM:tw3}),  
is given by 
${\sf S}_{\bf H} = 
\sqrt{\!{\bf \Lambda}_{\bf H}} \, {\bf O} \, \sqrt{\mathbf{H}^{-1}}$.
%
From (\ref{Ex:DiagH}), one writes
\begin{equation}
\sqrt{\mathbf{H}^{-1}} = {\sf O}'\sqrt{{\bf D}^{-1}}{\sf O}'^\top, 
\end{equation}
and it remains to determine the matrix $\bf O$ from the solution of
Eq.(\ref{SM:tw4}), which for the present case is  
\begin{equation}
{\bf O} \sqrt{\bf H} \, \mathsf J \, \sqrt{\bf H}\, 
{\bf O}^\top = {\bf \Lambda}_{\bf H}\mathsf J. 
\end{equation}
Using again (\ref{Ex:DiagH}) and the fact that ${\sf O}' \in {\rm Sp}(2n,\mathbb R)$,
then above equation becomes
\begin{equation}
{\bf O} {\sf O}'\sqrt{\bf D} \, \mathsf J \, \sqrt{\bf D}\, {\sf O}'^\top
{\bf O}^\top = {\bf O} {\sf O}' 
{\bf \Lambda}_{\bf H} \mathsf J \, {\bf O}'^\top
{\bf O}^\top  =
{\bf \Lambda}_{\bf H} \mathsf J,   
\end{equation}
thus ${\bf O}  = {\sf O}'^\top$ and the matrix ${\sf S}_{\bf H}$ becomes  
\begin{equation} \label{Eq:SympMatEnd}                                                     
{\sf S}_{\bf H} = 
\sqrt{\!{\bf \Lambda}_{\bf H}} \, \sqrt{\mathbf{D}^{-1}}\,{\sf O}'^\top, 
\end{equation}
for 
${\bf \Lambda}_{\bf H}$ in (\ref{Ex:SympEigH}), 
$\mathbf{D}$ in (\ref{Ex:SpecH}) and ${\sf O}'$ in (\ref{Ex:DiagH2}). 

With above matrix, the evolution of the normal mode coordinates is (\ref{Eq:NmFlux})
for $x'_\star = 0$ and ${\mathsf S}'_t$ in (\ref{Eq:NmEvol}) with ${\bf \Lambda}_{\bf H}$ 
in (\ref{Ex:SympEigH}). 
The thermal equilibrium state (\ref{Eq:thstate}) for the system described by the 
Hamiltonian in (\ref{Ex:QuantHam}) can be written in terms of creation-annihilation 
operators using the symplectic change of variables in (\ref{Eq:NmTransf}), 
with ${\sf S}_{\bf H}$ in (\ref{Eq:SympMatEnd}), $\xi = 0$, 
and $\mathbf W$ in (\ref{Eq:CompSymp}) for $n=3$. 
The resulting expression is Eq.(\ref{Eq:ThstateWT}) for $n=3$.

\subsection{Thermal State and Uncertainty Principle}
As learnt in Sec.\ref{Sec:WtSI}, 
the uncertainty relation when written in terms of symplectic eigenvalues 
(of the covariance matrix) is a structural property of the system and is independent 
of an operator-basis choice. 
For a Thermal state described by (\ref{Eq:ThstateWT}), it is convenient  
to write the uncertainty relation (\ref{Eq:UR}) in terms of the
creation-annihilation operators defined in (\ref{Eq:zdef}). 

To this end, it is opportune to deal
with the eigenvectors of the Hamiltonian 
$\hbar \mu_j (\hat a_j^\dag \hat a_j + \tfrac{1}{2})$, 
which are Fock sta\-tes\textsuperscript{\citenum{SMQuantMech,SMStatPhys}} 
denoted by $|\nu_j\rangle$ for $\nu_j = 0,..., \infty$; 
an eigenstate of the whole system is the tensor product state 
$ |\nu_1, ..., \nu_n\rangle := 
|\nu_1\rangle \otimes ...\otimes |\nu_n\rangle$.
Consequently, 
the mean value of a generic operator $\hat A$ is calculated through  
\begin{equation*}
\begin{aligned}
\langle \hat A \rangle &= {\rm Tr}(\hat \rho_{\rm T} \hat A )  
=\sum_{{\nu}_1 = 1}^{\infty} \!\!...\!\! \sum_{{\nu}_n = 1}^{\infty}
\langle \nu_1, ..., \nu_n | 
\hat\rho_{\rm T}\hat A |\nu_1, ..., \nu_n  \rangle.  
\end{aligned}
\end{equation*}
%
%

Defining $\Delta \hat z := \hat z - \langle \hat z \rangle$, 
see Sec.\ref{Sec:UP} of the main text, 
the covariance matrix  
\begin{equation} \label{Eq:CMZdef}
\tilde{\bf V}_{\!jk} = \tfrac{1}{2} {\rm Tr}\left[ 
\{\Delta\hat z_j, \Delta\hat z_k\} \hat \rho_{\rm T} \right] 
\end{equation}
for the thermal state in (\ref{Eq:ThstateWT}) 
is determined by calculating the following quantities:
\begin{equation}\label{SM:MVz}
\begin{aligned}
&\langle \hat a_j \rangle = \langle \hat a_j^\dagger \rangle = 0, \,\,\, 
\langle \hat a_j \hat a_k \rangle = 
\langle \hat a_j^\dagger a_k^\dagger \rangle = 0, \\ 
&\langle \hat a_j \hat a_k^\dagger\rangle  = 
\langle \hat a_j^\dagger \hat a_k \rangle + \delta_{jk}, \,\,\, 
\langle \hat a_j^\dagger \hat a_k \rangle = 
\langle \hat a_j^\dagger \hat a_j \rangle \delta_{jk}, 
\end{aligned}
\end{equation}
and
%
\begin{equation*}
\begin{aligned}
\langle \hat a_j^\dagger \hat a_j \rangle & = 
\sum_{\nu_j = 0}^\infty 
\langle \nu_j | \hat\rho_{\rm T}^{(j)}  
\hat a_j^\dagger \hat a_j| \nu_j \rangle =
\sum_{\nu_j = 0}^\infty\frac{ 
\nu_j {\rm e}^{-\beta \hbar \mu_j(\nu_j + \frac{1}{2}) }}
{ \frac{1}{2} {\rm csch}\left(\tfrac{1}{2} \beta\hbar \mu_j\right) } \\
& = \tfrac{1}{2} {\rm e}^{-\frac{1}{2}\beta \hbar \mu_j } 
{\rm csch}\left(\tfrac{1}{2} \beta\hbar \mu_j\right) 
= \left( {\rm e}^{\beta \hbar \mu_j } - 1 \right)^{-1}. 
\end{aligned}
\end{equation*}
Collecting all these mean-values into $\tilde{\bf V}$, 
see Eq.(\ref{Eq:zdef}), one finds
\begin{equation}\label{Eq:Covz}
\tilde{\bf V} = 
\frac{i \hbar}{2} 
\left( \begin{array}{cc}
{\bf 0}_n & \tilde{\bf N} \\  
\tilde{\bf N}  & {\bf 0}_n 
\end{array} \right)\!,   \,\, 
\end{equation}
where $\tilde{\bf N} := 2 \, {\rm Diag}(
\bar{\nu}_1,...,\bar{\nu}_n) + {\bf I}_n$ and 
\begin{equation} \label{Eq:BosOc}
\bar{\nu}_j := \langle \hat a_j^\dagger \hat a_j \rangle = 
\left( {\rm e}^{\beta \hbar \mu_j } - 1 \right)^{-1} \ge 0
\end{equation}
is called the bosonic occupation number\textsuperscript{\citenum{SMStatPhys}}.

Once the covariance matrix is obtained for the operators $\hat z$, 
it remains to write it for $\hat x$ 
through the transformation (\ref{Eq:NmTransf}). 
First note that, from Eq.(\ref{SM:MVz}), $\langle \hat z \rangle =  0$ and thus  
$ \langle \hat x \rangle = - {\bf H}^{-1} \xi$;
consequently $\Delta \hat z =  {\bf W}{\mathsf S}_{\bf H}^{-\top} \Delta \hat x$. 
Inserting this last relation into the definition (\ref{Eq:CMZdef}), 
similarly to (\ref{Eq:CovV}), one attains 
\begin{equation}\label{Eq:CovCRRel}
\tilde{\bf V} = 
{\bf W}\,{\mathsf S}_{\bf H}^{-\top} {\bf V} \,{\mathsf S}_{\bf H}^{-1}{\bf W}.
\end{equation}
It is essential to note that, 
while $\tilde{\bf V}$ in (\ref{Eq:CovCRRel}) 
is calculated with the thermal state written as in (\ref{Eq:ThstateWT}),
matrix ${\bf V}$ should be calculated with the thermal state written 
for the quadratic Hamiltonian $\hat h = H(\hat x)$. 
This is a mere consequence of the fact that the Hamiltonian
is subjected to the same transformation, see Eq.(\ref{Eq:DiagHam}), 
as it should be. 

Departing from the uncertainty relation (\ref{Eq:UR}), using Eq. (\ref{Eq:CovCRRel}),
and the fact that ${\mathsf S}_{\bf H}$ is symplectic, 
the uncertainty relation becomes\textsuperscript{\citenum{SMNoteTransf}}%
\begin{equation*}
\begin{aligned}
{\mathsf S}_{\bf H}^{\top}  {\mathbf W}^\ast 
\tilde{\bf V}   {\mathbf W}^\ast {\mathsf S}_{\bf H}
+ \frac{i\hbar}{2} {\sf J} \ge 0 &\Longleftrightarrow 
  {\mathbf W}^\ast 
\tilde{\bf V}   {\mathbf W}^\ast 
+ \frac{i\hbar}{2} {\sf J} \ge 0 \\ 
&\Longleftrightarrow
\left( \begin{array}{cc}
\tilde{\bf N} &  i {\bf I}_n\\  
 -i {\bf I}_n & \tilde{\bf N} 
\end{array} \right) \ge 0,
\end{aligned}
\end{equation*}
where Eq.(\ref{Eq:Covz}) was employed. 
The Euclidean eigenvalues of the last matrix are given by 
$\lambda_j^{\pm} = 2\bar{\nu}_j + 1 \pm 1,  j=1,...,n$, 
which are all non-negative, since $\bar{\nu}_j \ge 0$, 
see Eq.(\ref{Eq:BosOc}). 
In conclusion, every positive-definite quadratic Hamiltonian generates 
a genuine physical thermal state.   

By the end, note that since $\tilde{\bf V}$ is complex, it is not suitable 
for the Williamson theorem. 
However, it is still possible to determine the symplectic eigenvalues for the 
appropriate covariance matrix, which is $\bf V$. 
Writing explicitly $\bf W$, see Eq.(\ref{Eq:CompSymp}), in Eq.(\ref{Eq:CovCRRel}), 
one reaches
\begin{equation}
{\bf V} = 
\tfrac{\hbar}{2}  {\sf S}_{\bf H}^{\top} (\tilde{\bf N} \oplus \tilde{\bf N}) 
{\sf S}_{\bf H}.  
\end{equation}
However, the symplectic spectrum is invariant under 
a symplectic congruence, in such a way $
{\bf \Lambda}_{\bf V} = \tfrac{\hbar}{2} (\tilde{\bf N} \oplus \tilde{\bf N})$, 
thus the symplectic eigenvalues of the covariance matrix ${\bf V}$ are
$\mu'_j = \tfrac{\hbar}{2}(2 \bar{\nu}_j + 1), \, j=1,...,n$. 
Due to the definition of $\bar{\nu}_j$ in Eq.(\ref{Eq:BosOc}), 
the relation between the symplectic spectra of the Hamiltonian and 
the covariance matrix is 
\begin{equation}
{\bf \Lambda}_{\bf V} = \tfrac{\hbar}{2} {\rm coth}
\!\left( \tfrac{1}{2} \beta\hbar {\bf \Lambda}_{\bf H}\right),
\end{equation}
which is valid for any positive-definite quadratic Hamiltonian.



\begin{thebibliography}{99}          
\bibitem{arnold}
V.I. Arnol’d,
{\it Mathematical Methods of Classical Mechanics},  
Graduate Texts in Mathematics, 2nd ed. 
(Springer-Verlag, New York, 1989); 
\bibitem{landau1}
L.D. Landau and E.M. Lifshitz, {\it Mechanics}, 
Course of Theoretical Physics Vol.1, 3rd ed. 
(Elsevier, Oxford , 2005).
\bibitem{goldstein}
H. Goldstein, C.P. Poole Jr., and J.L. Safko, 
{\it Classical Mechanics}, 3rd ed. 
(Addison Wesley, London, 2000);
\bibitem{lemos}
N.A. Lemos, {\it Analytical Mechanics}, 
(Cambridge University Press, Cambridge, 2018).
\bibitem{gossonbook2006}
M. de Gosson, 
{\it Symplectic Geometry and Quantum Mechanics}, 
series Operator Theory: Advances and Applications 
(Birkh\"auser, Basel, 2006), Vol.166. 
\bibitem{littlejohn1986}
R.G. Littlejohn, 
``The Semiclassical Evolution of Wave Packets,'' 
Phys. Rep. {\bf 138}(4-5), 193--291 (1986). 
\bibitem{SupMatDem}
See Sec.SM1 of the Supplementary Material at [URL will be inserted by AIPP].
\bibitem{SupMatLag}
See Sec.SM3 of the Supplementary Material at [URL will be inserted by AIPP].
\bibitem{sakurai}
J.J. Sakurai and J. Napolitano,
{\it Modern Quantum Mechanics} 2nd Ed.
(Addison-Wesley, Boston, 2011).  
\bibitem{ballentine}
L.E. Ballentine, 
{\it Quantum Mechanics -- A Modern Development} 
(World Scientific, Singapore, 2000);
\bibitem{cohen}
 C.C.-Tannoudji, B. Diu, and F. Lalo{\"e}, 
{\it Quantum Mechanics} 
Vol 1: Basic Concepts, Tools, and Applications, 2nd ed.
(Wiley-VCH, Singapore, 2019).
\bibitem{landau2} 
L.D. Landau and E.M. Lifshitz, 
{\it Statistical Physics} Part 1, 
Course of Theoretical Physics Vol.5, 3rd ed. 
(Pergamon Press, Oxford, 1980);  
\bibitem{huang}
K. Huang, 
{\it Statistical Mechanics} 2nd ed., 
(John Wiley \& Sons, New York, 1987);  
\bibitem{pathria}
R.K. Pathria and P.D. Beale, 
{\it Statistical Mechanics} 3rd ed., 
(Elsevier Science, Amsterdam, 2011).
\bibitem{simon1994}
R. Simon, N. Mukunda, and  B. Dutta, 
``Quantum-noise matrix for multimode systems: 
     $U(n)$ invariance, squeezing, and normal forms,''
Phys. Rev. A {\bf 49}(3), 1567--1583 (1994).
\bibitem{SupMatEx}
Sec.SM5 of the Supplementary Material at [URL will be inserted by AIPP] 
contains three examples of physical systems where the application 
of the Williansom theorem is performed: 
Interacting Trapped Ions, Quantized Electromagnetic Field, 
and Thermal State and Uncertainty Principle. 
\bibitem{SupMatEx2}
See Sec.SM5-A of the Supplementary Material at [URL will be inserted by AIPP].
\bibitem{horn2013} 
R.A. Horn and C.R. Johnson, 
{\it Matrix Analysis} 2nd ed., 
(Cambridge University Press, New York, 2013). 
\bibitem{Note:InerDef}
The inertia of matrix is the triple containing 
the number of positive, negative, and null 
eigenvalues counting multiplicities, see for instance, Definition 4.5.6 
in Ref. \onlinecite{horn2013}, p.281. 
\bibitem{Note:InerLaw}
Theorem 4.5.8 in Ref. \onlinecite{horn2013}, p.252. 
\bibitem{silvester}
J.R. Silvester, 
``Determinants of Block Matrices,'' 
The Mathematical Gazette {\bf 84}(501), 460-467 (2000). 
\bibitem{Note:DetSimp}
There are several ways to show that, 
for a symplectic matrix $\sf S$, 
$\det {\sf S} = + 1$, but none of them is trivial. 
%
The most economical way is to use the concept 
of Pfaffian, for the definition see 
R. Vein and P. Dale, 
{\it Determinants and their Applications in Mathematical Physics} 
(Springer-Verlag, New York, 1999), pp.73-78.  
%
Denoting $\rm Pf$ as the Pfaffian of a matrix, 
${\rm Pf}({\sf S}^\top {\sf J} {\sf S}) = 
 (\det{\sf S}) {\rm Pf}({\sf J})$ and 
${\rm Pf}({\sf S}^\top {\sf J} {\sf S}) = 
{\rm Pf}({\sf J})$, 
and thus $\det{\sf S} = + 1$, 
as shown by Gosson (Ref. \onlinecite{gossonbook2006}, p.29).  
%
Arnol'd mechanics book (Ref. \onlinecite{arnold}, p.222) 
shows it using symplectic forms.
%
More recently, 
a longer proof but using only 
basic linear algebra was developed in 
D. Rim, 
``An elementary proof that symplectic matrices have determinant one,''
Adv. Dyn. Sys. Appl. {\bf 12}, 15--20 (2017).   
\bibitem{simon1999}
R. Simon, S. Chaturvedi, and V. Srinivasan, 
``Congruences and canonical forms for a positive matrix: Application to the
Schweinler–Wigner extremum principle,'' 
J. Math. Phys. {\bf 40}, 3632--3642 (1999). 
\bibitem{Note:Cor} 
Part (b) of Corollary 2.5.11 in Ref. \onlinecite{horn2013} (p.136),
using the notation of this work, is the following: 
%
Let $\tilde{\bf M} \in {\rm M}(m,\mathbb R)$. 
Then ${\tilde{\bf M} = -\tilde{\bf M}^\top}$ if and only if there 
is a real orthogonal ${\bf Q}$ and a non-negative integer $k$ 
such that 
${\bf Q} \tilde {\bf M} {\bf Q}^\top$ has the form
%
${\bf 0}_{m-2k} \oplus 
\omega_1 \left(\begin{smallmatrix}
                0 & 1\\
               -1 & 0
\end{smallmatrix}\right)
%
\oplus ... \oplus 
\omega_k\left(\begin{smallmatrix}
              0 & 1\\
              -1 & 0
              \end{smallmatrix}\right)$, 
with all $\omega_j > 0$.  
%
The case analyzed in this work has $m = 2n$ 
and $\det \tilde{\bf M} \ne 0$, consequently $\tilde{\bf M}$ is even dimensional 
and does not have null eigenvalues, 
thus $k = n$. Under these conditions, 
above canonical form becomes 
%
$\omega_1 \left(\begin{smallmatrix}
                0 & 1\\
                -1 & 0
               \end{smallmatrix}\right)
%
\oplus ... \oplus 
%
\omega_n \left(\begin{smallmatrix}
                0 & 1\\
                -1 & 0
               \end{smallmatrix}\right) $, 
%
which by a permutation of columns turns to 
${\mathsf J} ({\bf \Omega} \oplus {\bf \Omega})$ with 
${\bf \Omega} = {\rm Diag}(\omega_1,...,\omega_n)$.  
\bibitem{gosson2}
M.A de Gosson, 
``The symplectic egg in classical and quantum mechanics,''
Am. J. Phys. {\bf 81}(5), 328--337 (2013). 
\bibitem{SupMatQH}
The solution for a generic quadratic Hamiltonian is in 
Sec.SM2 of the Supplementary Material at [URL will be inserted by AIPP]. 
\bibitem{arnoldODE}
V.I. Arnol'd, 
{\it Ordinary Differential Equations}, 
(Springer-Verlag, New York, 1992).
\bibitem{Note:Anom}
This is not an anomalous behavior of the 
Hamiltonian scenario, 
see example 5.2 in N.A. Lemos (Ref. \onlinecite{lemos}, pp. 152-153) 
where the stability of the system depends 
on the fourth-order term of the Lagrangian.   
\bibitem{Note:VecOp}
To avoid misunderstandings, 
the operator $\hat q_j$ is the position operator related to the 
$j$\textsuperscript{th} degree of freedom and is a short notation to 
$\hat 1_{1}   \otimes \hat 1_{2} \otimes ...\otimes 
 \hat 1_{j-1} \otimes \hat q_j   \otimes 
 \hat 1_{j+1} \otimes ... \otimes \hat 1_{n}$, 
where $\hat 1_{j}$ is the identity operator on the Hilbert space associated to 
the $j$\textsuperscript{th} degree of freedom. 
%
The same consideration applies to momenta operators. 
\bibitem{Note:VeOp2}
%
For a real vector $\eta:= (\eta_1,...,\eta_{2n}) \in {\mathbb R}^{2n}$, 
the sum $\hat x'= \hat x + \eta$ should be interpreted as an 
operator vector with components 
$\hat x'_j = \hat x_j + \eta_j \hat 1_j$. 
%
For a matrix ${\bf A} \in {\rm M}(2n)$,  
$\hat x' = {\bf A } \hat x$ is a vector with components 
$\hat x'_j = \sum_{k = 1}^{2n} {\bf A }_{jk} \hat x_k$ for  $j =1 ,..., 2n$. 
%
\bibitem{Note:QuantProb}
In the scope of this work, it is enough to consider a canonical 
symmetric quantization,
which consists in replacing products $\hat x_j\hat x_k$ by 
its symmetric version $1/2(\hat x_j\hat x_k+\hat x_k\hat x_j)$.
%
Note that, incidentally, $\mathbf H^\top = \mathbf H$ 
implies a symmetric quantization for the classical Hamiltonian in 
Eq.~(\ref{Eq:ClQuadLinHam}).
%
The quantization of a classical system is itself an open problem 
of quantum mechanics,  
see S.T. Ali and M. Engli\v{s}, 
``Quantization Methods: A Guide for Physicists and Analysts,''
Rev. Math. Phys. {\bf 17}(4), 391-490 (2005).
\bibitem{NegHeatCap}
F. Staniscia, A. Turchi, D. Fanelli, P.H. Chavanis, and G. De Ninno, 
``Negative Specific Heat in the Canonical Statistical Ensemble,'' 
Phys. Rev. Lett. {\bf 105}, 010601--010605 (2010); 
%
H.A. Posch, H. Narnhofer, and W. Thirring,
``Dynamics of unstable systems,'' 
Phys. Rev. A {\bf 42}, 1880--1890 (1990).
\bibitem{nicacio16}
F. Nicacio, 
``Weyl–Wigner representation of canonical equilibrium states,'' 
J. Phys. A: Math. Theor. {\bf 54}, 055004, 1--30 (2021).
\bibitem{Schro}
E. Schrödinger, {\it Zum Heisenbergschen Unschärfeprinzip}
Phy\-si\-ka\-lisch-mathematische Klasse XIX, p.296-303 (1930).
English translation:  
{\it About Heisenberg Uncertainty Relation}
\href{http://arxiv.org/abs/quant-ph/9903100}{arXiv:quant-ph/990300 v3 (2008)} by 
A. Angelow and M.C. Batoni.
\bibitem{Rob}
H.P. Robertson,
``The Uncertainty Principle,'' 
Phys. Rev. {\bf 34}, 163--164 (1929).  
\bibitem{Note:NumIneq}
For $j,k \in \{1,...,n\}$, 
there will be a set of $n(2n +1)$ inequalities Eq.~(\ref{Eq:RobUR}) 
composed by 
(i) $n$ inequalities for  
$\hat A = \hat q_k$ and $\hat B = \hat q_k$; 
%
(ii) $(n^2-n)/2$ inequalities raised by pairs 
$(\hat A = \hat q_j, \hat B =\hat q_k)$ with $j \ne k$  
since $[\Delta \hat q_j,\Delta \hat q_k] = 0$; 
%
(iii) $n^2$ inequalities from pairs like
$(\hat q_j,\hat p_k)$, 
which will be the same inequalities as 
the ones generated by the pairs 
$(\hat p_k,\hat q_j)$;  
%
(iv) $(n^2+n)/2$ pairs of the form 
$(\hat p_j,\hat p_k)$ generates inequalities 
which are counted as the previous 
$(\hat q_j,\hat q_k)$. 
%
Note that the total number of inequalities is 
the number of independent elements 
of a symmetric matrix in ${\rm M}(2n,\mathbb R)$.   
\bibitem{narcowich}
F.J. Narcowich, ``Geometry and uncertainty,'' 
J. Math Phys. {\bf 31}, 354--364 (1990).
\bibitem{Note:Notation} 
To clarify the notation, 
observe that for a real vector 
$y \in {\mathbb R}^n$, 
the object $yy^\top$ 
is a $n\times n$ real matrix with elements 
$(yy^\top)_{jk} = y_j y_k$. 
%
In the same sense,
$\Delta\hat x \Delta\hat x^\top$ 
is a $2n \times 2n$ matrix with elements 
$\Delta\hat x_j \Delta\hat x_k$ and its mean-value, 
$\langle \Delta\hat x \Delta\hat x^\top \rangle$, 
has elements 
$\langle \Delta\hat x \Delta\hat x^\top \rangle_{jk} = 
\langle \Delta\hat x_j \Delta\hat x_k \rangle$.
\bibitem{SupMatPosit}
See Sec.SM4 of the Supplementary Material at [URL will be inserted by AIPP].
\bibitem{williamson}
J. Williamson, 
``On the Algebraic Problem Concerning the Normal Forms of Linear Dynamical Systems,'' 
Amer. J. Math. {\bf 58}, 141--163 (1936).
\bibitem{bogoliubov}
N.N. Bogoliubov, 
{\it On a new method in the theory of superconductivity}, 
Nuovo Cim {\bf 7}, 794–805 (1958). 
\bibitem{simon2000}
R. Simon, 
``Peres-Horodecki Separability Criterion for Continuous Variable Systems,''
Phys. Rev. Lett. {\bf 84}(12), 2726--2729 (2000). 
\bibitem{adesso} 
G. Adesso and F. Illuminati, 
``Entanglement in con\-ti\-nuous-variable systems: 
recent advances and current perspectives,''
J. Phys. A \textbf{40}(28), 7821--7880 (2007).
\bibitem{nicacio8}
F. Nicacio and F.L. Semião, 
``Coupled harmonic systems as quantum buses in thermal environments,'' 
J. Phys A: Math. Theor. {\bf 49}(37), 375303-1 -- 375303-30 (2016);
%
{\it Id}, ``Transport of correlations in a harmonic chain,'' 
Phys. Rev. A, {\bf 94}, 012327-1 -- 012327-12 (2016).
\end{thebibliography}

\begin{thebibliography}{99}          
\bibitem{SMClassMec}
V.I. Arnol’d,
{\it Mathematical Methods of Classical Mechanics},  
Graduate Texts in Mathematics, 2nd ed. 
(Springer-Verlag, New York, 1989); 
%
L.D. Landau \& E.M. Lifshitz, {\it Mechanics} 
(Volume 1 of {\it Course of Theoretical Physics}, 
Elsevier, Oxford 3rd Ed, 2005);
%
H. Goldstein, C.P. Poole Jr. \& J.L. Safko, 
{\it Classical Mechanics}
(Addison Wesley, London, 3rd Ed., 2000);
%
N.A. Lemos, {\it Analytical Mechanics}, 
(Oxford University Press, Cambridge, 2018).
\bibitem{SMarnoldODE}
V.I. Arnol'd, 
{\it Ordinary Differential Equations}, 
(Springer-Ver\-lag, New York, 1992).
\bibitem{SMPaulTrap}
W. Paul, 
{\it Electromagnetic traps for charged and neutral particles}, 
\href{https://doi.org/10.1103/RevModPhys.62.531}
{Reviews of Modern Physics {\bf 62}, 531 (1990)};
%
P.K. Ghosh, {\it Ion Traps} 
(Oxford University Press, New York, 1995).  
\bibitem{SMleibfried}
D. Leibfried \& R. Blatt, C. Monroe, and D. Wineland, 
{\it Quantum dynamics of single trapped ions},
\href{https://doi.org/10.1103/RevModPhys.75.281}
     {Rev. Mod. Phys. {\bf 75}, 281 (2003)}.
\bibitem{SMzoller}
J.I. Cirac \& P. Zoller, 
{\it Quantum Computations with Cold Trap\-ped Ions}, 
\href{https://doi.org/10.1103/physrevlett.74.4091}
     {Physical Review Letters {\bf 74}, 4091 (1995).}
\bibitem{SMnicacio3}
F. Nicacio, K. Furuya, \& F.L. Semião,
{\it Motional entanglement with trapped ions and a nanomechanical resonator},  
\href{https://doi.org/10.1103/PhysRevA.88.022330}{Physical Review A {\bf 88}, 022330 (2013)};
\href{https://arxiv.org/pdf/1212.0711.pdf}{arXiv:1212.0711 [quant-ph] (2013)}.
\bibitem{SMNoteUniq}
If $\det{\bf U} \ne 0$, 
the solution for $q_\star$ 
is unique and given by 
$q_\star = q_0 -  \tfrac{C}{d}{\bf U}               
\left(\begin{smallmatrix}
            1 \\
            -1 
      \end{smallmatrix} \right)$; 
otherwise, there can be multiple 
equilibrium points $q_\star$.
\bibitem{SMNotePointT} 
{Implementing the point transformation 
$q' = q - q_\star$ in (\ref{Ex:QuadLag}), 
this Lagragian attains (\ref{SM:Lagragian})
with ${\bf T} = m {\bf I}_{2}$. 
%
The potential matrix in (\ref{SM:LagDiag}) is thus 
$\tilde{\bf U} = m^{-1} \bf U$ and 
$\tilde {\bf O}$ is the diagonalizing matrix of $\bf U$.} 
\bibitem{SMNoteSqueez}
Squeezing is a property related to the variances 
of measurements, see Eq.(\ref{Eq:variance}). 
%
Suppose that there is a quantum state such that 
$\langle \Delta \hat q_j^2  \rangle 
 \langle \Delta \hat p_j^2  \rangle = c$, 
where of course $c \ge \hbar^2/4$, 
see Eq.(\ref{Eq:HeisUR1D}). 
%
A new state is said squeezed with respect to 
the former if one of the variances 
is increased while the other is decreased 
by the same factor,  
%
that is, the new variances are such that 
$\langle \Delta \hat q_j'^2  \rangle = 
s^2 \langle \Delta \hat q_j^2 \rangle$ and
%
$\langle \Delta \hat p_j'^2  \rangle = 
s^{-2} \langle \Delta \hat p_j^2 \rangle$. 
%
In a squeezed state, measurements of 
one variable will have a sharper distribution, 
%
while the one for the conjugate variable will 
be broader, however their product 
is left unchanged, 
$\langle \Delta \hat q_j'^2  \rangle 
 \langle \Delta \hat p_j'^2  \rangle = c$. 
%
For the quantum electromagnetic field, 
position and momentum are called 
quadrature\textsuperscript{\citenum{SMQuantOptics}}  and are 
identified by relations 
(\ref{Eq:zdef}) and (\ref{Eq:ComplRep}). 
%
The mentioned squeezing effect 
is generated by a Hamiltonian like 
$\hat H_1$ in Eq.(\ref{Ex:QuantHam}), 
while $\hat H_2$ generates the same effect 
but taking into account different fields \textsuperscript{\citenum{SMQuantOptics}}. 
\bibitem{SMQuantOptics}
M. Scully \& M. Zubairy, 
{\it Quantum Optics} (Cambridge University Press, Cambridge, 1997);
%
D.F. Walls \& G.J. Milburn,
{\it Quantum Optics} (Springer-Verlag, Berlin, 2nd ed. 2008);
%
W.P. Schleich, 
{\it Quantum Optics in Phase Space} (Wiley‐VCH Verlag, Berlin, 2001);
%
G. Grynberg, A. Aspect \& C. Fabre, 
{\it Introduction to Quantum Optics --- 
From the Semi-classical Approach to Quantized Light} 
(Cambridge University Press, Cambridge, 2010).
\bibitem{SMnicacio8}
F. Nicacio \& F.L. Semião, 
{\it Coupled harmonic systems as quantum buses in thermal environments}, 
\href{https://doi.org/10.1088/1751-8113/49/37/375303}
     {Journal of Physics A {\bf 49}, 375303 (2016)};
\href{https://arxiv.org/pdf/1601.07528.pdf}{arXiv:1601.07528 [quant-ph](2016)}.
\bibitem{SMQuantMech}
J.J. Sakurai \& J. Napolitano,
{\it Modern Quantum Mechanics}
(Addison-Wesley, Boston, 2nd Ed. 2011).  
L.E. Ballentine, 
{\it Quantum Mechanics -- A Modern Development} 
(World Scientific, Singapore 2000);
 C.C.-Tannoudji, B. Diu \& F. Lalo{\"e}, 
{\it Quantum Mechanics}
(Wiley-VCH, Singapore, 2005).
\bibitem{SMStatPhys} 
L.D. Landau \& E.M. Lifshitz, 
{\it Statistical Physics} Part 1, 
(Volume 5 of {\it Course of Theoretical Physics}, 
Pergamon Press, Oxford 3rd Ed, 1980);  
%
K. Huang, 
{\it Statistical Mechanics} 
(John Wiley \& Sons, 2$^\text{nd}$ Ed. 1987);  
%
R.K. Pathria \& P.D. Beale, 
{\it Statistical Mechanics} (Elsevier Science, 1996).
\bibitem{SMNoteTransf}
The transformation (\ref{Eq:CovCRRel}) 
is not a congruence between $\bf V$ and $\tilde{\bf V}$ due to 
the matrices $\bf W$. 
By the same reason, it is not possible to ensure that
${\bf W} {\bf \Delta} {\bf W}$ is a positive-definite matrix, 
which forbids the statement
${\bf \Delta} \ge 0 
\Longleftrightarrow {\bf W} {\bf \Delta} {\bf W} \ge 0$, 
about (\ref{Eq:UR}).
\end{thebibliography}
\end{document}